\documentclass[iop,twocolappendix,floatfix]{emulateapj}

\usepackage{lineno}

\usepackage{epsfig}
\usepackage{apjfonts}
\usepackage{amsmath,color,bm}
\usepackage{booktabs}
\usepackage[breaklinks,colorlinks,urlcolor=blue,citecolor=blue,linkcolor=blue]{hyperref}
\usepackage[utf8]{inputenc}
\usepackage[T1]{fontenc}
\usepackage[normalem]{ulem}

\shorttitle{Fast and efficient Bayesian method to search for strongly lensed gravitational waves}
\shortauthors{Barsode et al}

\usepackage{natbib}
\usepackage{nameref}
\usepackage{graphicx}
\usepackage{graphics}
\usepackage[space]{grffile}
\usepackage{latexsym}
\usepackage{amsfonts,amsmath,amssymb}
\usepackage{url}
\usepackage[utf8]{inputenc}
\usepackage{fancyref}
\usepackage{hyperref}
\usepackage{xcolor}
\usepackage[normalem]{ulem}

\def\blu{{\mathcal{B}^L_U}}
\def\rlu{{\mathcal{R}^L_U}}

\def\slu{{\mathcal{S}^L_U}}
\def\plu{{\mathcal{P}^L_U}}
\def\bo{{\mathcal{B}_\mathrm{eq}}}
\def\b{{\mathcal{B}}}
\def\theq{{\theta_\mathrm{eq}}}
\def\thb{{\theta_\mathrm{b}}}
\def\thbI{{\theta_\mathrm{b1}}}
\def\thbII{{\theta_\mathrm{b2}}}
\def\dthb{{\Delta\theta_\mathrm{b}}}
\def\dphi{{\Delta\phi}}
\def\dt{{\Delta t}}
\def\rom{\mathrm{m}}
\def\mur{{\mu_\mathrm{r}}}
\def\HL{{\mathcal{H}_L}}
\def\HU{{\mathcal{H}_U}}
\def\mgal{{\mathcal{M}_\mathrm{gal}}}
\def\fapcat{{\mathrm{FAP}_\mathrm{cat}}}
\def\catsig{{\sigma_\mathrm{cat}}}
\def\fappair{{\mathrm{FAP}_\mathrm{pair}}}
\def\pe{{\textsc{pe}}}

\begin{document}

\title{Fast and efficient Bayesian method to search for strongly lensed gravitational waves}

\author{Ankur Barsode$^{1\ a}$}
\email{$^a$ankur.barsode@icts.res.in}
\author{Srashti Goyal$^{2,1\ b}$}
\email{$^b$srashti.goyal@aei.mpg.de}
\author{Parameswaran Ajith$^{1,3\ c}$}
\email{$^c$ajith@icts.res.in}

\affiliation{$^1$~International Centre for Theoretical Sciences, Tata Institute of Fundamental Research, Bangalore 560089, India}
\affiliation{$^2$~Max Planck Institute for Gravitational Physics (Albert Einstein Institute), Am Mühlenberg 1, D-14476 Potsdam-Golm, Germany}
\affiliation{$^3$~Canadian Institute for Advanced Research, CIFAR Azrieli Global Scholar, MaRS Centre, West Tower, 661 University Ave., Suite 505, Toronto, ON M5G 1M1, Canada}

\begin{abstract}

A small fraction of the gravitational-wave (GW) signals from binary black holes observable by ground-based detectors will be strongly lensed by intervening objects such as galaxies and clusters. Strong lensing will produce nearly identical copies of the GW signals separated in time. These lensed signals must be identified against a background of unlensed pairs GW events, some of which may appear similar by accident. This is usually done using fast, but approximate methods that, for example, check for the overlap between the posterior distributions of a subset of binary parameters, or using slow, but accurate joint Bayesian parameter estimation. In this work, we present a modified version of the posterior overlap method dubbed ``PO2.0" that is mathematically equivalent to joint parameter estimation while still remaining fast. We achieve a significant gain in efficiency by incorporating informative priors about the binary and lensing populations, selection effects, and all the inferred parameters of the binary. For binary black hole signals lensed by galaxies, our improved method can detect 65\% lensed events at a pair-wise false alarm probability of $\sim 2\times 10^{-6}$. Consequently, we have a $13\%$ probability of detecting a strongly lensed event above $2.25\sigma$ significance during 18 months of observation by the LIGO-Virgo detectors at their current sensitivity. We also show how we can compute the joint posteriors of the lens and source parameters from a pair of lensed events by reweighting the posteriors of individual events in a computationally inexpensive way.
\end{abstract}


\section{Introduction}
\label{sec:intro}

Gravitational lensing of gravitation waves (GWs) occurs when the gravity of a massive ``lens'' lying along the line of sight alters the path of GWs. If the gravitational size of the lensing object  ($\sim$ $G M_\mathrm{lens}/c^2$) is much larger than the wavelength of GWs, the phenomenon of lensing can be treated under the geometric optics approximation. In strong lensing, lensed GWs take multiple paths before refocusing on the observer, therefore appearing as multiple copies of the same waveform. These copies are identical to the source waveform except for an overall magnification, an overall phase shift of $n\pi/2$ (where the integer $n$ is known as the Morse factor \citep{dai2017waveforms}), and an overall time delay. From the observer's point of view, these appear as nearly identical signals arriving from the same patch of sky, but having different apparent luminosity distances and arrival times, with their coalescence phases differing by $n\pi/2$.

Because the magnification, time delay, and Morse phase depend on the mass profile of the lensing object, strongly lensed GWs offer direct probes of collapsed structure in the universe, with applications ranging from measuring the velocity dispersion in galaxies \citep{xu2022please} to studying the nature of dark matter \citep{jana2024probing}, and even for constraining cosmological parameters \citep{jana2024strong}. Strongly lensed images may also be used for localizing the source with greater precision than that allowed by individual detections \citep{hannuksela2020localizing, wempe2024detection}. They could also be used to test the polarisation content of GWs more precisely since additional images effectively multiply the number of detectors~\citep{Goyal:2020bkm}.

Detection of such magnified images in the data may be used to search for additional, fainter images \citep{ng2024uncovering}, or wave optics effects \citep{seo2022improving}. Even the non-observation of strongly lensed images \citep{hannuksela2019search, mcisaac2020search, LIGOScientific:2021izm, abbott2023search, janquart2023follow} can constrain compact dark matter \citep{barsode2024constraints} and binary black hole (BBH) formation channels \citep{leong2024constraining}.

All of these applications of strong lensing of GWs are conditioned on our ability to identify such lensed pairs of signals in the first place, or confidently rule them out. Though lensing does not alter GW polarisations, the observed lensed GW signals may appear different. One reason for this is that the Earth rotates between the arrival of different images, causing the antenna patterns to change, which results in those images having different combinations of the GW polarisations (see, e.g.,~\cite{Goyal:2020bkm}). The detector noise will also bias the lensed images differently. Even more serious is the possibility of false positives whereby two independent, unlensed signals may appear consistent purely by accident. The problem of lensing identification then becomes a classification problem of distinguishing between truly lensed image pairs from unlensed event pairs~\citep{haris2018identifying}, which can be significantly limited by false alarms~\citep{ccalicskan2023lensing}.

We expect $\sim 0.1-1\%$ of all events detectable by ground-based GW detectors to be strongly lensed by galaxies and clusters \citep{xu2022please, wierda2021beyond}. Since several hundred GW detections are expected in the next few years~\citep{abbott2020prospects}, the first detection of strongly lensed GWs is imminent. However, identifying a strongly lensed pair of signals among millions of possible pairs in a catalog of GW signals is challenging. Apart from the large computational cost, the task is made harder by the presence of false positives where unrelated GW signals resemble each other purely by chance. The probability of such false alarms grows in proportion to the number of pairs detected, making the problem of lensing identification a two-fold challenge in computational cost and false alarm probability, both of which grow quadratically with the number of detections.

A computationally efficient method for identifying strongly lensed image pairs was developed by \cite{haris2018identifying}, where they derived the Bayesian likelihood ratio (Bayes factor) between the lensed and unlensed hypotheses. They showed that the Bayes factor can be written as an inner product of the posterior distributions of the source parameters estimated from two signals, inversely weighted by the prior. Intuitively, this makes sense: Since strongly lensed signals have many identical parameters (such as the detector frame masses, spins, sky location, etc.), posterior distributions of such parameters should show significant overlap. The inverse weighting by the prior takes care of the accidental overlap of the posteriors due to prior choices. The posteriors were marginalized over parameters (e.g., luminosity distance) that would be biased between the two lensed images. They further combined this Bayes factor with another likelihood ratio (between the lensed and unlensed hypotheses) based on the time delay between the two signals. This makes use of the fact that time delays between lensed images follow a different distribution than that between randomly arriving unlensed signals. This combined statistic was able to identify 80\% lensed image pairs with a false alarm probability $\sim 10^{-5}$ in the Advanced LIGO-Virgo network \citep{aasi2015advanced, acernese2014advanced}.

Alternate approaches have followed, with the aim of reducing computational costs using faster approximations to the posterior distributions \citep{goyal2024rapid}, or to rapidly check waveform consistency using machine learning \citep{goyal2021rapid, magare2024slick}, or cross-correlation of time series data \citep{chakraborty2024glance}. Search pipelines have also been developed to look for sub-threshold images that may be a counterpart of super-threshold detections \citep{li2023targeted, li2023tesla, goyal2024rapid}. Methods have also been developed with the aim of easily weeding out false alarms, for example, by checking the \textit{in}consistency between quantities that are better measured by the detectors \citep{ezquiaga2023identifying}. \cite{more2022improved} proposed to enhance the time delay likelihood ratio by including information of the relative magnification of the two images. We shall elaborate more upon some of these methods in the subsequent sections as we relate and compare them with our statistic.

\cite{liu2021}, followed by \cite{lo2023bayesian} presented a fully Bayesian formalism for the strong lensing odds ratio by performing joint parameter estimation (PE) over common and lensing-biased parameters of the GW signals. Their formalism also took into consideration the binary population, as well as selection effects. However, such a joint PE was computationally expensive. An alternate approach by \cite{janquart2021fast} took advantage of the parameter constraints obtained by unlensed PE to achieve faster convergence, and it was later optimized in \cite{janquart2023return}. Even then, joint PE can take up to an hour to compute the statistic for a single pair, making it difficult to study the background and estimate false alarm probabilities.

In this work, we improve the posterior overlap method to capture prior information of source and lens populations, as well as to fold in the information stored in parameters biased by lensing. Our new statistic, which we call ``PO2.0'', offers the promise of potentially reaching the same efficiency as joint PE, but with evaluation times of a few minutes. We believe this should be possible because the two signals' individual posterior distributions in principle contain all the information about their joint distribution as well, provided their noises are uncorrelated and we use the correct prior.

We benchmark the performance of PO2.0 at identifying galaxy-lensed BBHs from an unlensed background with false alarm probabilities as low as $2\times 10^{-6}$, or, equivalently, with significance level as high as $\leq 2.25 \sigma$ for an 18 month observing period of the LIGO-Virgo detectors at the O4 run sensitivity \citep{H1L1V1-psd-O3O4O5}. This is enabled both by the low evaluation times of our method, as well as by the aid of fast PE code \texttt{cogwheel} \citep{roulet2022removing, islam2022factorized} which allows us to simulate the background of our Bayesian model selection at a low computational cost.

The rest of the paper is organized as follows. In \S\ref{sec:BLU_PO2} we outline the derivation of the PO2.0 statistic and show how it reduces to various other strong lensing search statistics proposed in the literature. In \S\ref{sec:ROCs} we benchmark its performance in terms of Receiver Operating Characteristic (ROC) curves, highlighting the contribution to efficiency from the use of informative priors, selection effects, and biased parameters. In the same section, we also compare the performance to some of the other ``fast'' statistics that exist in the literature. \S\ref{sec:catsig} discusses the prospects of detecting and confidently identifying a strongly lensed GW event in a realistic catalog. We show how PO2.0 can be used to measure the parameters of a lensed event in \S\ref{sec:po2_pe}. Finally, in \S\ref{sec:conclusion} we conclude our paper with a discussion about future plans and extensions for PO2.0.

\section{The PO2.0 statistic}
\label{sec:BLU_PO2}
In this section, we derive our new Bayesian statistic that incorporates all the available information about the binary system, assuming only that the two signals are non-overlapping and the noise contained in them is uncorrelated. We follow it up by showing how PO2.0 is related to other lensing identification methods described in the literature. In particular, we show that PO2.0 is mathematically equivalent to joint PE while being much faster to evaluate.

A strongly lensed binary merger appears as multiple copies of the same source waveform polarisations $h^{+,\times}(f; \theta)$, where $\theta$ denotes the parameters of the binary (component masses, spins, sky location, etc.). The polarisations of the $j$-{th} image, $h_j$, differs from $h$ by an overall magnification $\mu_j$, a time delay $\dt_j$, and a constant phase shift $\dphi_j$ (also known as Morse phase) of $0, \pi/2$ or $\pi$ \citep{dai2017waveforms}
\begin{equation}
{h}^{+,\times}_j\left(f ; \theta, \mu_j, \dt_j, \dphi_j\right)= {\left|\mu_j\right|^{1/2}} h^{+,\times}\left(f ; \theta\right)~  e^{i \, (2 \pi f \dt_j + \dphi_j)}
\end{equation}
The magnification changes the overall amplitude of the signal and is therefore degenerate with the luminosity distance to the source. Similarly, the time delay and phase shift are degenerate with the arrival time and coalescence phase of the binary\footnote{In case of $\dphi=\pi/2$ (``type II'' images), the degeneracy between coalescence phase and Morse phase can be broken if there is significant contribution from higher order modes of gravitational wave radiation \citep{dai2020search, vijaykumar2023detection, janquart2021identification}, but this is usually a weak measurement so we ignore it in this study.}. Consequently, when two strongly lensed images are observed, one can typically measure only the \textit{relative} magnification, time delay, and phase difference. At each GW detector, one can then write the observed polarizations $h^{+,\times}(f,\theta)$ of the second image in terms of those of the first as
\begin{eqnarray}
h^{+,\times}_2(f;\theq, \thbII) & = & h^{+,\times}_1(f;\theq, \thbII=\thbI +  \dthb) \nonumber \\
& \equiv & {\mur}^{1/2} ~ h^{+,\times}_1(f;\theq, \thbI) ~ e^{i \, (2 \pi f \Delta t + \Delta \phi)}
\label{eq:lensd_wave_h2}
\end{eqnarray}
where,
\begin{itemize}
\item $\theq$ are the subset of $\theta$ that remain the same between strongly lensed copies (i.e., detector frame masses, spins, sky location, inclination, and polarization angles).
\item $\thbI$ and $\thbII$ are the remaining parameters in $\theta$ that change due to strong lensing and may be different between the two images viz. luminosity distance $d_L$, coalescence phase $\phi_c$, and arrival time $t_c$. Note that these are different from the source's true $d_L, \phi_c$, and $t_c$ since they are biased by the magnification, Morse phase, and time delay of each image.
\item $\dthb = \thbII - \thbI$\footnote{here, to make the notation compact, we overload the symbols $+$ and $-$ to also mean $d_{L,2} = d_{L,1} / \sqrt{\mur} \equiv d_{L,1}\ ``+"\ \mur$ and $\mur = (d_{L,1} / d_{L,2})^2 \equiv d_{L,2}\ ``-"\ d_{L,1}$.} are the biases introduced by strong lensing. These consist of relative magnification $\mur \equiv (d_{L,1} / d_{L,2})^2$, Morse phase difference $\dphi \equiv \phi_{c,2} - \phi_{c,1}$ and relative time delay $\dt \equiv t_{c,2}-t_{c,1}$. We choose to denote the first signal that arrives at the detectors as image 1 so that $\dt > 0$.
\end{itemize}

Our aim is to express a detection statistic (essentially a function of the data) in terms of these variables and simplify it into a tractable formula that can be evaluated over a given pair of GW signals. Similar to \cite{haris2018identifying}, we adopt the Bayes factor $\blu$ as our detection statistic, defined as the ratio of likelihoods of getting the observed data $\{d_1, d_2\}$ under the lensed ($\HL$) and unlensed ($\HU$) hypotheses\footnote{Throughout the paper, we use the same symbol $P$ to denote different probability distributions, distinguishing between them using only their arguments. We often refer to only a subset of dimensions of a joint distribution, which is to be understood as a distribution that is marginalized over the other parameters.}
\begin{equation}
\blu = \dfrac{P(d_1,d_2\mid \HL)}{P(d_1,d_2\mid \HU)}.
\end{equation}
The marginal likelihoods (Bayesian evidences) of the hypotheses can be obtained by marginalizing over the parameters of the model. Assuming that the two signals are non-overlapping, and their noise realizations are uncorrelated, this yields 

\begin{widetext}
\begin{equation}
\blu  = \frac{1}{\displaystyle\prod\limits_{j=1}^2 \int d\theta\ P(d_j \mid \theta) ~~ P(\theta \mid \HU)}  \int d\theq\ d\thbI\ d\dthb ~~ P(d_1 \mid \theq, \thbI) ~~ P(d_2 \mid \theq, \thbII=\thbI + \dthb) ~~ P(\theq, \thbI, \dthb \mid \HL) \vphantom{\int}.
\label{eq:lensing_likelihood}
\end{equation}

We want to compute this Bayes factor using posteriors of the binary parameters obtained from $d_1$ and $d_2$ where the PE is performed without considering any lensing effects. The advantage is that this will allow us to bypass the need to evaluate the lensing likelihood $P(d_1, d_2 \mid \HL)$ from all the signal pairs using a computationally expensive joint PE. We show in Appendix~\ref{sec:BLU_derivation} that $\blu$ can be calculated by marginalizing appropriately reweighted products of posteriors over the signal parameters, leading to the PO2.0 statistic 
\\
\begin{equation}
\label{eq:BLU_expression}
\blu = \frac{1}{\displaystyle\prod\limits_{j=1}^2  \int d\theta\ \frac{P(\theta \mid d_j)} {P_{\textsc{pe},j}(\theta)} ~~ {P(\theta \mid \HU)}}  ~~
\int d\theq\ d\thbI\ d\dthb ~~ \dfrac{P(\theq, \thbI \mid d_1)}{P_\textsc{PE,1}(\theq, \thbI)}~~ \dfrac{P(\theq, \thbII=\thbI + \dthb \mid d_2)}{P_\textsc{PE,2}(\theq, \thbII=\thbI + \dthb)} ~~ P(\theq, \thbI, \dthb \mid \HL).
\end{equation}
\end{widetext}

Above, ${P_{\textsc{pe},j}(\theta)}$ denotes the prior used in the PE of individual signals while $P(\theta \mid \HU)$ denotes the appropriate astrophysical prior under $\HU$. Also,  $P(\theq, \thbI \mid d_1)$ denotes the posterior on $\theq$ and $\thbI$ obtained from signal 1, while $P(\theq, \thbII=\thbI + \dthb \mid d_2)$ denotes the posterior on $\theq$ and $\thbII$ obtained from signal 2, evaluated at $\thbII = \thbI + \dthb$. Finally, $P(\theq, \thbI, \dthb \mid \HL)$ is the astrophysical prior on the parameters $\theq, \thbI, \dthb$ under $\HL$.

\subsection{Comparison with other search methods}

The inner product presented in eq.~\ref{eq:BLU_expression} is a generalization of the posterior overlap statistic from~\cite{haris2018identifying}.
\begin{equation}
\label{eq:Bo_haris_et_al}
\bo = \int d\theq\ \dfrac{P(\theq \mid d_1) ~ P(\theq \mid d_2)}{P(\theq)}.
\end{equation}
This can be derived from eq.~\ref{eq:BLU_expression} by making two simplifying assumptions:
\begin{enumerate}
\item Consider only the posteriors on $\theq$, by marginalising over $\thb_j$:
\begin{equation}
P(\theq \mid d_j) = \int d\thb_j ~ P(\theq, \thb_j \mid d_j).
\label{eq:marg_post}
\end{equation}
\item Assume that the astrophysical priors under $\HL$ and $\HU$ are identical to the PE priors:
\begin{equation}
P(\theq \mid \HL) = P(\theq \mid \HU) = P_{\textsc{pe},j}(\theq).
\label{eq:prior_assum_haris}
\end{equation}
\end{enumerate}

Simpler approximations to the above have been explored in the literature \citep{goyal2024rapid},  by keeping only a subset of $\theq$ (such as the chirp mass and sky location) in the integral after approximating the posteriors by a Gaussian function and assuming uniform priors\footnote{Note that the Bhattacharya coefficient \citep{bhattacharyya1946measure}  used by \cite{goyal2024rapid} uses the square root of the posteriors inside the integral in eq.~\ref{eq:Bo_haris_et_al}. This will lead to a different ranking for the lensed pairs. \cite{goyal2024rapid} found that eq.~\ref{eq:Bo_haris_et_al} was more efficient in identifying lensed true lensed events than the Bhattacharya coefficient.}.

\cite{cheung2023mitigating} pointed out that \cite{haris2018identifying}'s assumption of population priors being identical to the PE priors was unrealistic, and proposed an improved statistic that avoided this oversimplification.
\begin{equation}
\label{eq:Bo_cheung_et_al}
\bo^\mathrm{(C)} = \frac{\int d\theq\ {\frac{P(\theq \mid d_1)}{P_\textsc{PE,1}(\theq)} ~ \frac{P(\theq \mid d_2)}{P_\textsc{PE,2}(\theq)} ~ P(\theq | \HU)}}{ \prod\limits_{j=1}^2  \int d\theq\ \frac{P(\theq \mid d_j)} {P_{\textsc{pe},j}(\theq)} ~ {P(\theq \mid \HU)} } .
\end{equation}
This can be derived from eq.~\ref{eq:BLU_expression} by making two simplifying assumptions:
\begin{enumerate}
\item Consider only the posteriors on $\theq$, by marginalising over $\thb_j$ (same as eq.~\ref{eq:marg_post}).
\item Assume same  astrophysical priors under $\HL$ and $\HU$, but different from PE priors. That is, instead of eq.~\ref{eq:prior_assum_haris}, assume
\begin{equation}
P(\theq \mid \HL) = P(\theq \mid \HU) \neq P_{\textsc{pe},j}(\theq).
\label{eq:prior_assum_cheung}
\end{equation}
\end{enumerate}
Note that their assumption of the \textit{same} population prior for lensed and unlensed events is not entirely correct (Fig.\ref{fig:population_m1_dL}).

\cite{dai2020search} used a statistic for incorporating extrinsic and lensing parameters in their eq.~3, which is similar to our eq.~\ref{eq:BLU_expression}. In fact, their two statistics -- eqs. 1 and 3 in their paper -- are factored out versions of our eq.~5, with the caveat that they did not use the population information in the BBH parameter priors, nor did they account for potential correlations between some of the intrinsic, extrinsic, and lensing parameters in the individual signal posteriors or joint priors.

Finally, joint PE methods are aimed at computing the Bayes factor described in eq.~\ref{eq:lensing_likelihood}. Their precise expressions (e.g., \cite{liu2021identifying}'s eq.~5, \cite{lo2023bayesian}'s eq.~17, \cite{janquart2021fast}'s eqs.~7, 10) differ from our eq.~\ref{eq:BLU_expression} in the following ways:
\begin{enumerate}
\item We use the detectable population as our prior whereas they use the intrinsic population. They incorporate selection effects by multiplying the evidence ratio by a constant factor, which we do not need\footnote{A constant multiplicative factor leaves the false positive and true positive probabilities unaffected. However, false positives are more likely in the parameter space favored by the population of \textit{detectable} events, as opposed to that favored by the \textit{intrinsic} population, and one should use the former for down-ranking false positives more efficiently \citep{cheung2023mitigating}.}.
\item We use a slightly different parameterization in terms of $\theq, \thbI, \dthb$ instead of $\theq, \thbI, \thbII$
\item Through the lensed population prior, we allow \emph{all} the parameters to be correlated with each other.
\end{enumerate}
All of these differences are minor, though formally, they may slightly improve the distinguishing power of our statistic. The key difference, however, lies in the evaluation method: while joint PE methods calculate the joint likelihood under the lensed hypothesis by sampling over BBH and lensing parameters, we calculate it \textit{back} from the individual posterior distributions of the two signals' parameters. This is enabled by splitting the joint likelihood in the specific form shown in eq.~\ref{eq:likelihood_factorization}. Our $\blu$ can therefore be computed with far fewer computational resources, allowing us to do deep background studies necessary for robust identification of lensed events.

We can show that (refer to Appendix~\ref{sec:BLU_simplification}), in eq.~\ref{eq:BLU_expression}, if one approximates the posterior of the arrival time $t_c$ with a delta function at the measured value, integrals over the parameters $t_{c,1}, \dt$ can be factorized into a ratio of prior distributions of the time delay evaluated at its measured value $\dt^\rom$. This is exactly the $\rlu$ statistic from \cite{haris2018identifying} that they multiply with parameter overlap to define a combined statistic $\bo \times \rlu$, where,
\begin{equation}
\rlu = \dfrac{P(\dt^\rom \mid \HL)}{P(\dt^\rom \mid \HU)}.
\end{equation}
While \cite{haris2018identifying} motivated its form in a heuristic fashion, we derive $\rlu$ naturally from the likelihood ratio eq.~\ref{eq:BLU_expression}, highlighting the link between time delay priors and parameter overlap.

This link can be extended to \cite{more2022improved}'s $\mgal$ statistic\footnote{the ``gal'' here stands for galaxies, the most dominant lenses in the context of strong lensing of GWs observable by ground-based detectors. In this paper, however, we use the same symbol $\mgal$ for all lens populations.}
if one makes a further approximation that the luminosity distance and coalescence phase posteriors are delta functions as well. One can then follow similar steps as those in Appendix~\ref{sec:BLU_simplification} to show that such an approximation leads to all 6 of $\thbI$ and $\dthb$ factorizing out into a ratio of \textit{joint} lensed and unlensed priors of time delay $\Delta t$, magnification ratio $\mur$, and phase difference $\Delta \phi$ evaluated over their measured values $\dt^\rom, \mur^\rom, \dphi^\rom$
\begin{equation}
\label{eq:mgal}
\mgal = \dfrac{P(\dt^\rom, \mur^\rom, \dphi^\rom \mid \HL)}{P(\dt^\rom, \mur^\rom, \dphi^\rom \mid \HU)}
\end{equation}

In practice, the above expression is evaluated over a measure of central tendency, such as the mean or median of the posterior distributions. \cite{more2022improved} noted that $d_L$ and $\phi_c$ are in fact not as accurately measured as the time delay, but nevertheless, they expected this statistic to perform better than $\rlu$ because it includes additional information about magnification and Morse phase. We show in \S\ref{sec:ROCs} that, in reality, the biases introduced by wrongly approximating $d_L$ and $\phi_c$ posteriors as delta functions actually worsen the efficiency brought in by the information they contain. These biases are removed by our formalism by correctly folding in the measurement uncertainties in $d_L$ and $\phi_c$.

To summarise the discussion so far, we have shown that various lensing search statistics that work either by checking the consistency between two signals using an ``overlap'' in inferred parameters, or using ratios of informative priors, can be thought of as subsets or approximations of PO2.0, which itself is equivalent to joint PE.

There are also alternative statistics proposed in the literature that do not use the Bayesian overlap of posteriors for determining consistency of signals, but rather use other ``distance'' measures to quantify how much the inferred parameters differ. \cite{ezquiaga2023identifying} check for the \textit{inconsistency} between the phases measured at the individual detectors at a particular frequency since these are the better measured properties of a signal. This makes their pipeline effectively a vetoing pipeline than a detection pipeline.
Search techniques based on machine learning \citep{goyal2021rapid, magare2024slick} or cross-correlation of time series data \citep{chakraborty2024glance} work in a completely different way from all the posterior probability-based methods we have discussed so far, and we will not discuss them in this paper.

\subsection{Implementation}
\label{sec:priors}

\begin{figure*}[t]
\centering
\includegraphics[width=\columnwidth]{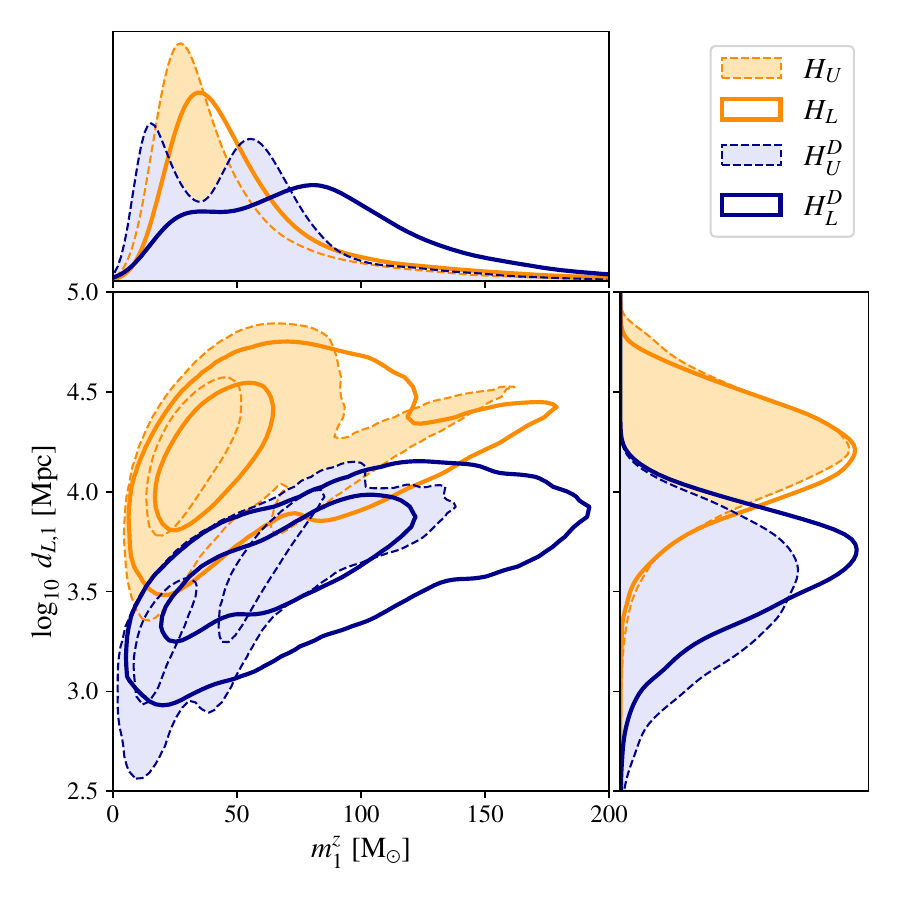}
\includegraphics[width=\columnwidth]{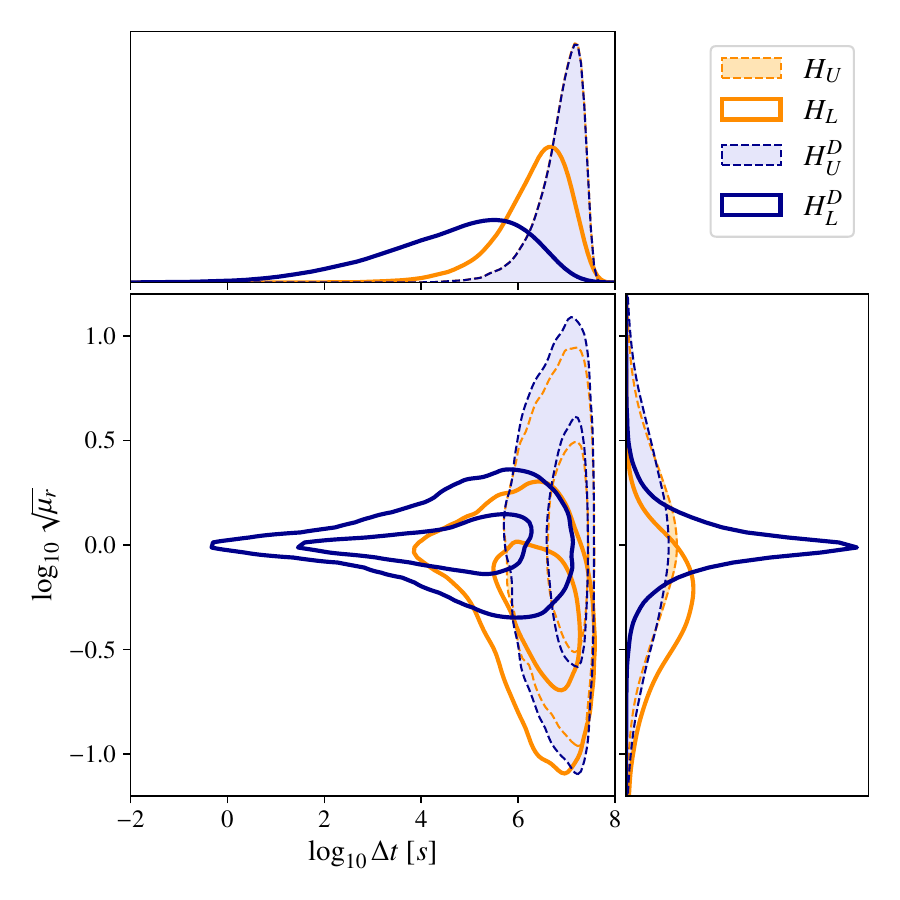}
\caption{\emph{Left panel}: The astrophysical prior $P(m_1^z, \log_{10}{d_{L,1}} \mid \mathcal{H})$ on the detector frame primary mass $m_1^z$ and the \textit{apparent} luminosity distance $d_{L,1} \equiv d_L/\sqrt{\mu_1} $ of the first image (or the only image in case of unlensed) under the lensed ($\mathcal{H} = \HL$) and unlensed ($\mathcal{H} = \HU$) hypotheses. The top and side panels show their marginalized distributions. The orange shades show the intrinsic population and the blue shades (with a superscript $D$ in the legend) show the detected population. These distributions are marginalized over all other parameters. \emph{Right panel:} Same as the left figure, except that the figure shows the prior $P(\log_{10}{\Delta t}, \log_{10}{\sqrt{\mur}} \mid \mathcal{H}) $ of the lensing time delay $\Delta t$ and the square root of the magnification ratio $\mur$.}
\label{fig:population_m1_dL}
\end{figure*}

In our implementation, we compute the kernel density estimates (KDEs) of posterior distributions $P(\theq, \thb \mid d_j)$ using the posterior samples produced by running PE codes on individual signals $d_j$.  We also sample $\dthb$ from the prior $P(\dthb \mid \HL)$ and re-evaluate $P(\theq, \thbII = \thbI + \dthb \mid d_2)$ and $P(\theq, \thbI, \dthb \mid \HL)$ to compute $\blu$  using eq.~\ref{eq:BLU_expression}. Even after factorizing out $\rlu$, this in general involves two 14-dimensional integrals over the binary parameters $\{\theq, \thbI\}$ (in the denominator -- excluding $t_c$), and a 16-dimensional integral over the binary parameters $\{\theq, \thbI\}$ as well as lensing biases $\dthb$ (in the numerator -- excluding $t_{c,1}$ and $\dt$). While it is possible to evaluate them using importance sampling methods, the result is too noisy due to the high dimensionality.

However, using our knowledge of the behavior of the BBH population as well as the GW likelihood function, we can simplify the high-dimensional integrals as products of independent low-dimensional integrals. It turns out that, if factorized properly, the loss in accuracy due to such approximation is well compensated by the gain in precision. In Appendix~\ref{sec:BLU_simplification}, we show how to simplify eq.~\ref{eq:BLU_expression} further by ignoring the PE priors and by factorizing sky overlap and Morse phase out of the integrals. Our numerical algorithm for evaluating that simplified expression is detailed in Appendix~\ref{sec:BLU_evaluation}.

Even after reducing the dimensionality, the KDEs could be prone to numerical errors. This problem can be alleviated to some extent by using more posterior samples at the cost of slower evaluation. Fortunately, we find that even with $\sim 10^4$ samples, errors in estimating the $\blu$ are similar to those encountered in stochastic sampling estimates (see Appendix~\ref{sec:BLU_evaluation}). At this precision, we can detect over 60\% of all lensed events at low false alarm probability ($\sim 2 \times 10^{-6}$). For the future, we are also exploring more accurate alternatives for reconstructing the posteriors that may help reduce the errors and improve the detection efficiency further.

The population priors $P(\theta \mid \HU)$ and $P(\theq, \thbI, \dthb \mid \HL)$ in eq.~\ref{eq:BLU_expression} are the joint distribution functions of detector frame parameters describing the BBH signal and lensing. We obtain these distributions by performing astrophysical simulations of BBH mergers, their lensing by galaxies, and by applying selection effects as applicable for LIGO-Virgo detectors at their O4 sensitivity. These are detailed in Appendix~\ref{sec:BBH_simulations}. Distributions of some of these parameters (marginalized over the rest) are shown in Fig.~\ref{fig:population_m1_dL}.

One can see that the distributions of unlensed and lensed events are qualitatively different. Lensed events tend to have higher redshifted masses since lensing is more likely at higher redshifts \citep{dai2017effect}. This also causes them to have higher luminosity distances, though lensing magnification may considerably change how the \textit{apparent} luminosity distance distribution looks in comparison to that for the unlensed events.

\section{Estimating the efficiency of PO2.0}
\label{sec:ROCs}
In this section, we evaluate the performance of PO2.0 for identifying strongly lensed GWs, making use of simulated GW signals. Since generating Bayesian posteriors of BBH parameters from a large number of simulated events is computationally expensive, we test PO2.0 on the dominant mode GW signals from quasicircular BBHs having aligned spins, for which PE can be performed at low computational cost using the \texttt{cogwheel} PE code~\citep{roulet2022removing, islam2022factorized}. We emphasize, however, that the method presented in the previous section is generally applicable to any pair of non-overlapping GW signals. The details of our simulation including injection dataset and PE are given in Appendix~\ref{sec:BBH_simulations}.

We adopt a frequentist approach for quantifying our confidence in identifying a lensed event by treating the Bayes factor as a ranking statistic. This is most easily visualized using an ROC curve, showing the detection efficiency as a function of false alarm probability. As outlined in Appendix~\ref{sec:BBH_simulations}, we use a simulated PE dataset of $\sim 10^3$ unlensed events, leading to $\sim 5 \times 10^5$ unlensed ``background'' pairs, allowing us to probe false alarm probabilities down to  $\sim 2\times 10^{-6}$. Our ``foreground'' simulation consists of $\sim 10^3$ truly lensed pairs, so our estimated efficiencies have a shot noise of $\sim 3\%$.

In this section, we talk about only the \textit{pairwise} false alarm probabilities, i.e. the probability of a single unlensed pair crossing a given $\blu$ threshold. The false alarm probabilities for a \textit{catalog} of GW signals will be discussed in the next section.

\subsection{Improvement due to population priors}
\begin{figure}[t]
\centering
\includegraphics[width=\columnwidth]{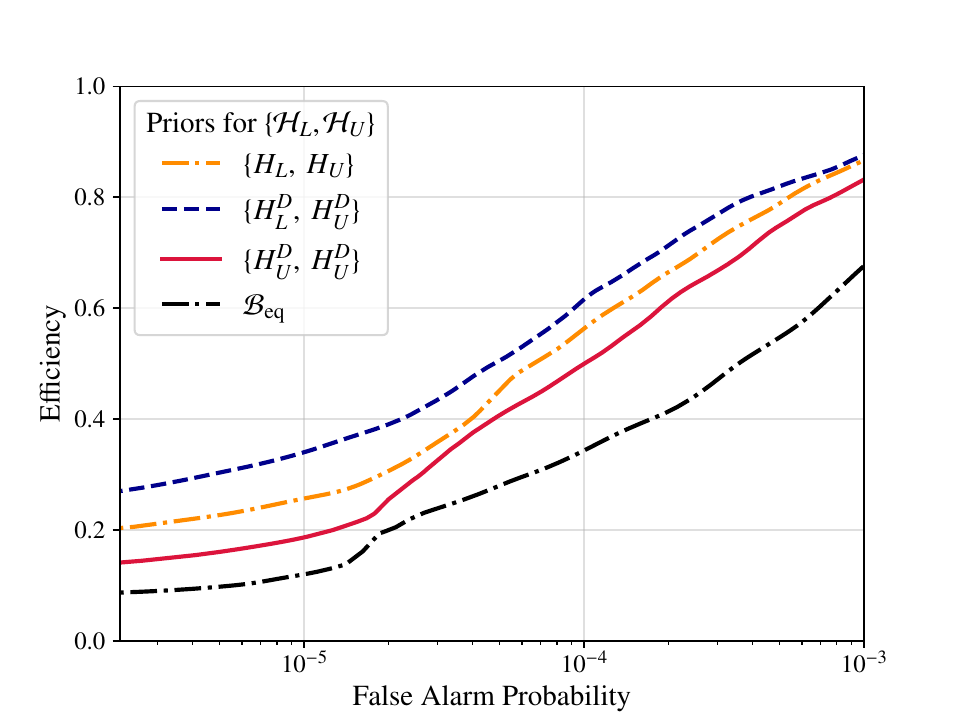}
\caption{ROC curves showing the improvement in detection efficiency due after incorporating population priors with ($H_L^D, H_U^D$) or without ($H_L, H_U$) selection effects. Only equal parameters $\theq$ are included in the calculation of $\blu$ in eq.~\ref{eq:BLU_expression}. Also shown in an ROC assuming that both unlensed as well as lensed events follow the same prior $H_U^D$. The ROC corresponding to \cite{haris2018identifying}'s model agnostic overlap statistic $\bo$ (eq.~\ref{eq:Bo_haris_et_al}) is also shown for reference.}
\label{fig:ROC_priors}
\end{figure}

We begin by discussing the improvement in detection efficiency due to the inclusion of population priors in computing the Bayes factor. We compute the $\blu$ in eq.~\ref{eq:BLU_expression} with different choices of population priors, with and without selection effects, but excluding $\thbI$ and $\dthb$. As we show in Fig.~\ref{fig:ROC_priors}, incorporating informative priors improves the detection efficiency at all false alarm probabilities. Furthermore, the improvement is higher if one includes selection effects in modeling the population, about $20\%$ at all false alarm probabilities. Therefore, we essentially verify \cite{cheung2023mitigating}'s assertion that one should use the population of detectable events when calculating lensing statistics to enhance their detection efficiency.

\cite{cheung2023mitigating} also noted that once all the binary parameters were included in calculating the $\blu$, the statistic may become too sensitive to the underlying population. We verify this hypothesis by using the distribution of \textit{unlensed} detectable signals $H_U^D$ as the \textit{lensed} population. As Fig.~\ref{fig:population_m1_dL} (left plot) shows, the mass distributions in these two populations are significantly different, so this serves as a good playground to test the effects of a ``wrong'' population model. The corresponding ROC is plotted in the same Fig.~\ref{fig:ROC_priors} and shows a clear degradation in efficiency due to the wrong population assumption. Since the $\blu$ calculated with population priors including selection effects performs best (apart from being the correct thing to do), we stick to using these priors for the rest of this paper.

\subsection{Improvement due to inclusion of $\thbI, \dthb$}
\begin{figure}[t]
\centering
\includegraphics[width=\columnwidth]{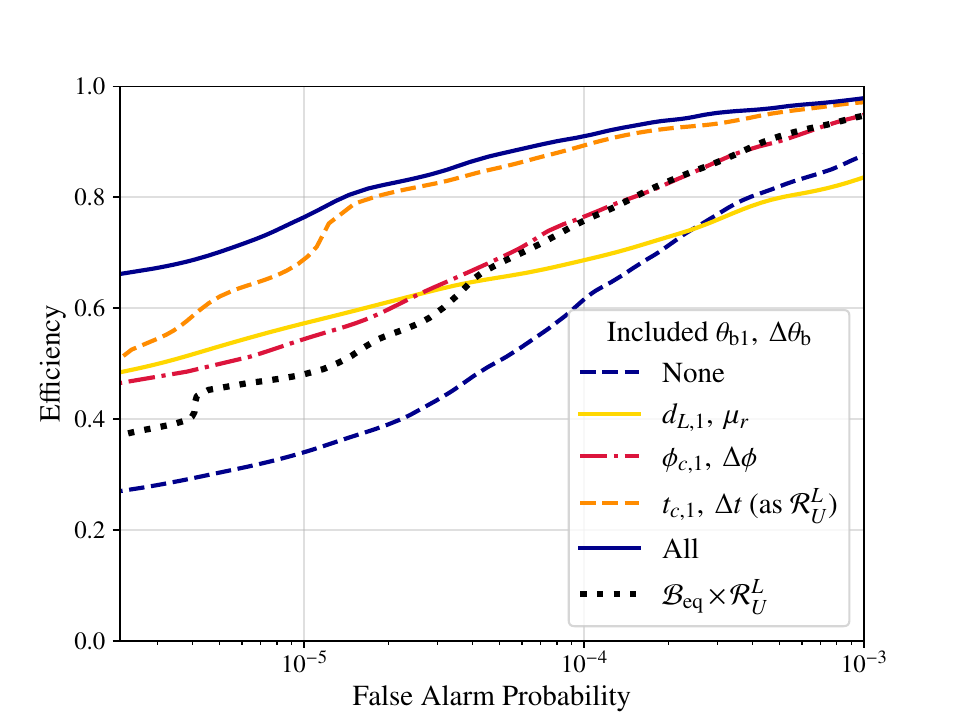}
\caption{ROC curves showing the improvement in detection efficiency due to the inclusion of biased parameters $\thbI$ and biases $\dthb$, assuming population priors including selection effects, $H_U^D$ and $H_L^D$. The equal parameters $\theq$ are included throughout. The ROC corresponding to \cite{haris2018identifying}'s detection statistic $\bo \times \rlu$, is also shown for reference.}
\label{fig:ROC_Det_biased_params}
\end{figure}

Next, we include the three pairs of additional parameters $\thbI, \dthb$. Here, $\thbI$ includes the luminosity distance, the coalescence phase, and the arrival time, while $\dthb$ includes the magnification ratio, the Morse phase, and the time delay. Figure~\ref{fig:ROC_Det_biased_params} shows that, individually, each of these pairs of parameters brings improvement in detection efficiency, especially at low false alarm probabilities. The most informative are the arrival time and time delay, included in the form of $\rlu$, as they add about 20-40\% to the detection efficiency.

The same figure shows that the detection efficiency improves further when one combines all six of these parameters with $\theq$ and uses the correct population priors with selection effects. This is the new posterior overlap statistic in its full glory and is about $65\%$ efficient in identifying lensed pairs at a false alarm probability of $\sim 2 \times 10^{-6}$. This is a $\sim70\%$ relative improvement over \cite{haris2018identifying}'s $\bo \times \rlu$ at low false alarm probabilities.

One may notice that the ROC curve corresponding to \cite{haris2018identifying}'s $\bo \times \rlu$ in Fig.~\ref{fig:ROC_Det_biased_params} has worse efficiency than that presented in the original paper's Fig.~9. For example, they claim to achieve $\sim 75-80\%$ efficiency at a false alarm probability of $10^{-5}$, whereas we find it to be only about $48\%$. The discrepancy is because, unlike our simulation (right panel of Fig.~\ref{fig:population_m1_dL}), \cite{haris2018identifying}'s lensing time delay distribution has very low support at larger values, which helps them achieve higher efficiency via the $\rlu$.

\subsection{Improvement due to different $\theq$}

Next, we factorize the $\theq$ parameters into sets of one or two $\theq$ at a time and evaluate eq.~\ref{eq:BLU_expression} for each set using 1D or 2D histograms. This exercise will help demonstrate how much different parameters contribute towards the distinguishing power of PO2.0 while simultaneously showing a -- less accurate -- but computationally efficient way of estimating the Bayes factor. Similar low-cost statistics were introduced in \cite{goyal2024rapid} for the mass overlap, though using a Gaussian approximation for the posteriors and without using any prior information from the BBH population.

\begin{figure}[t]
\centering
\includegraphics[width=\columnwidth]{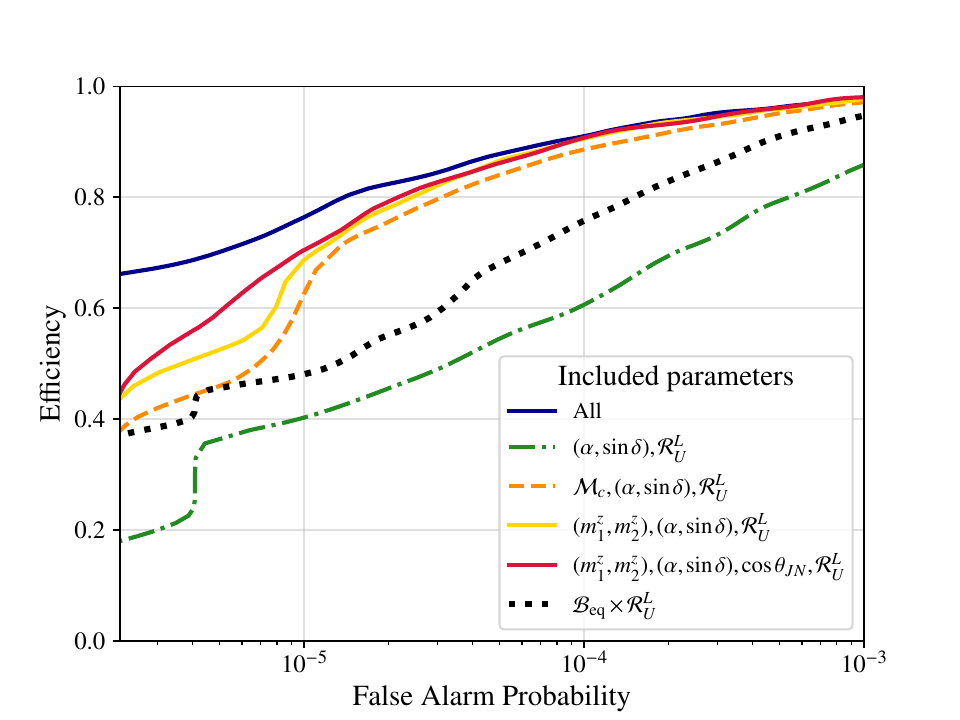}
\caption{ROC curves comparing the full version of PO2.0 with fast but approximate statistics computed by combining the overlaps in various parameter subsets such as sky localization $(\alpha, \sin\delta)$, chirp mass $\mathcal{M}_c$, component masses $(m_1^z, m_2^z)$ and inclination $\cos\theta_{JN}$. All statistics, except  \cite{haris2018identifying}'s $\bo \times \rlu$, are calculated assuming population priors including selection effects, ($H_U^D$ and $H_L^D$), and $\rlu$ is multiplied throughout.}
\label{fig:ROC_Det_hist_approx}
\end{figure}

Figure~\ref{fig:ROC_Det_hist_approx} shows that, when combined with $\rlu$, overlaps in only the chirp mass $\mathcal{M}_c$ and sky parameters $(\alpha, \sin\delta)$ together have better efficiency than just sky overlap, and even \cite{haris2018identifying}'s $\bo \times \rlu$. This efficiency is further enhanced if one also includes the mass ratio (by taking overlap in $(m_1^z, m_2^z)$ instead of $\mathcal{M}_c$) and the inclination angle $\cos\theta_{JN}$. We therefore conclude that, if limited by computational resources, combining overlaps in various 2D/1D subsets of the full parameter space -- $(m_1^z, m_2^z), (\alpha,\sin\delta)$ and $\cos\theta_{JN}$ -- together with $\rlu$, is a valid option for quick calculations. However, one should use the complete expression of the $\blu$ for better sensitivity, especially at low false alarm probabilities.

\subsection{Comparison with other low latency pipelines}
So far, we have been comparing our $\blu$ statistic with \cite{haris2018identifying}'s, on which it is based. Here, we compare PO2.0 with two other low latency lensing search pipelines proposed in the literature, namely $\mgal$ by \cite{more2022improved}, and \texttt{phazap} by \cite{ezquiaga2023identifying}.

\begin{figure}[t]
\centering
\includegraphics[width=\columnwidth]{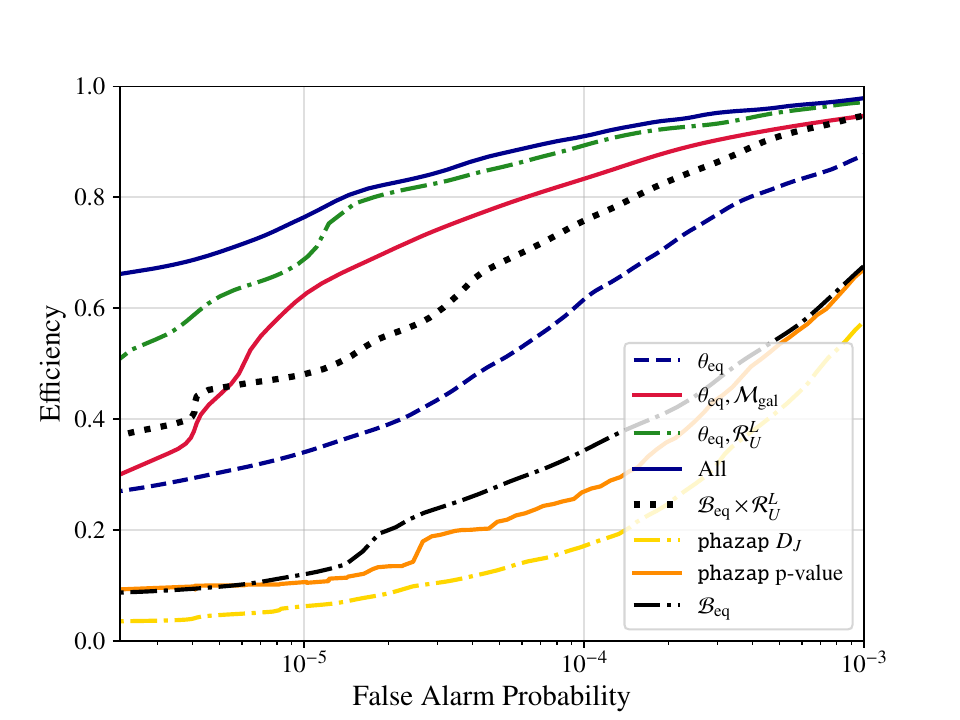}
\caption{Comparison of ROC curves of PO2.0 with \cite{more2022improved}'s $\mgal$, and \cite{ezquiaga2023identifying}'s \texttt{phazap}. The PO2.0 $\blu$'s were calculated assuming population priors including selection effects ($H_U^D$ and $H_L^D$), with parameters included as indicated by the legend.}
\label{fig:ROC_Mgal_Phazap}
\end{figure}

The $\mgal$ statistic \citep{more2022improved} was introduced as a drop-in replacement for $\rlu$ to include the information contained in all three $\dthb$. We find that, instead of using $\thbI, \dthb$ parameters in PO2.0, if we multiply it by $\mgal$, the improvement in efficiency is not that great (Fig.~\ref{fig:ROC_Mgal_Phazap}). In fact, it is even worse than the ROC obtained by multiplying posterior overlap in $\theq$ just by $\rlu$. This is because, unlike the arrival time $t_c$, posterior distributions of luminosity distance $d_L$ and coalescence phase $\phi_c$ are far from delta functions, and factorizing them out into $\rlu$-like ratio of magnification $\mur$ and Morse phase $\dphi$ priors can lead to large biases, degrading efficiency. PO2.0, on the other hand, correctly marginalizes over the uncertainties in each of these parameters, achieving a $10\%$ higher efficiency over $\rlu$, and $30\%$ higher efficiency over $\mgal$, at low false alarm probabilities.

We also plot the ROCs corresponding to \cite{ezquiaga2023identifying}'s \texttt{phazap}, a pipeline designed to veto out unlensed pairs using the inconsistency (quantified in terms of a ``distance'' $D_J$ and the related p-value) in their measured phases at detectors. Similar to \cite{haris2018identifying}'s $\bo$, this method does not include any prior information about populations and lensing. Our results in Fig.~\ref{fig:ROC_Mgal_Phazap} show that their pipeline is nearly as efficient at distinguishing between lensed and unlensed pairs as $\bo$ itself, though both of them are far less efficient than PO2.0 as it includes the population modeling of source and lenses.

\section{Detection in a catalog of GW signals}
\label{sec:catsig}
So far in this paper, we have discussed the efficiency of PO2.0 at identifying one pair of signals as lensed images of the same BBH event, or as two independent unlensed BBH events, based on the measured value of the statistic. However, in a real observing scenario, we may detect hundreds of GW signals, leading to tens of thousands of possible pairs. This significantly increases the probability of misclassifying an unlensed pair as lensed \citep{hannuksela2019search,ccalicskan2023lensing}, as the probability that \textit{at least} one unlensed pair in the catalog crosses the measured value of the statistic is much higher than that for a \textit{given} unlensed pair
\begin{equation}
\label{eq:fap_cat_definition}
\fapcat = 1 - (1 - \fappair)^{N_p} \approx N_p ~  \fappair
\end{equation}
where $\fappair$ is the \textit{pairwise} false alarm probability that we have been discussing so far, $\fapcat$ is the \textit{catalog} false alarm probability, and $N_p= N (N-1)/2$ is the total number of pairs one can form from a catalog of $N$ detected signals.

One can equivalently express the catalog false alarm probability in terms of a \emph{significance level} in the catalog
\begin{equation}
\catsig = \sqrt{2} ~ \texttt{erfc}^{-1}(\fapcat)
\end{equation}
where $\texttt{erfc}^{-1}$ is the inverse of the complementary Gaussian error function. Our simulation currently consists of $\sim 5\times 10^5$ unlensed pairs, letting us probe $\fappair \gtrsim 2 \times 10^{-6}$. This, together with our nominal estimate of $RT=150$ signals in O4 (for an observing duration of $T=18$ months and detection rate $R=100$ events per year\footnote{based on 40 GW detections in O4 with $P_{BBH}>50\%$ between 13th April 2024 and 5th September 2024. Data taken from \textsc{gracedb} \citep{gracedb-o4}}) leads to a maximum $\catsig=2.25$ that we can reliably probe in this study.

\begin{figure}[t]
\centering
\includegraphics[width=0.95\columnwidth]{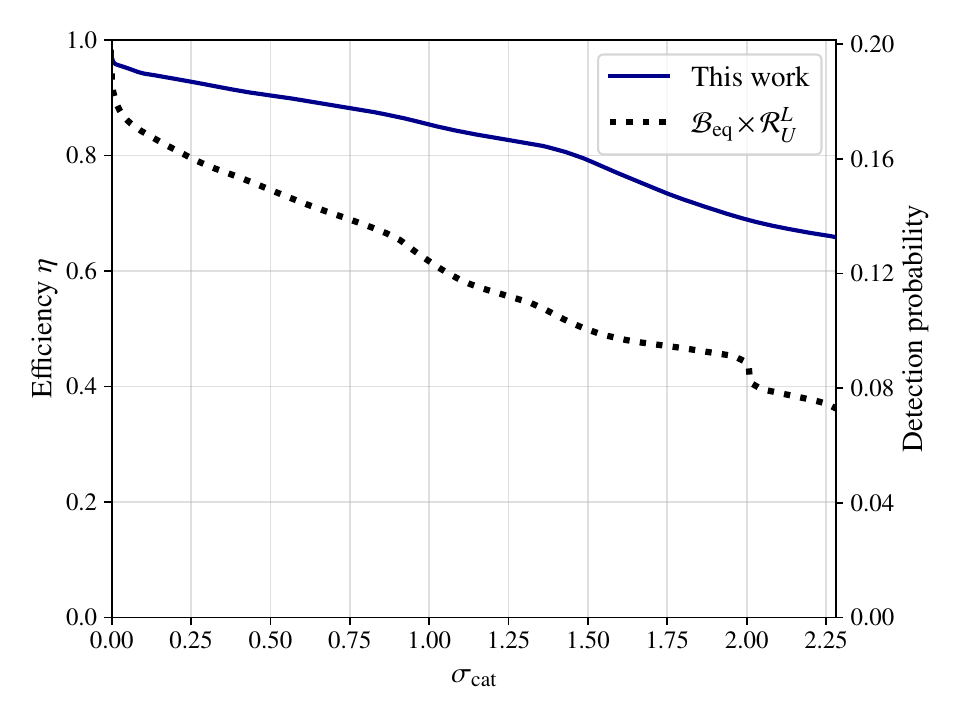}
\caption{Efficiency of detecting strongly lensed pairs above a catalog significance level $\catsig$. The second y-axis on the right shows the probability of detecting at least 1 lensed event in O4. The ROC curve corresponding to this work was calculated assuming population priors including selection effects, ($H_U^D$ and $H_L^D$), and all the available information in $\theq, \thbI, \dthb$.}
\label{fig:ROC_cat_sig}
\end{figure}

Using the nominal rates, we define the \emph{detection probability} as the probability of detecting at least 1 lensed pair of signals in a given catalog
\begin{equation}
\label{eq:pdet_definition}
\mathrm{detection\ probability} = 1 - e^{-u R T \eta}
\end{equation}
where $\eta$ is the detection efficiency that we have been discussing in the previous section, and $uRT$ is the Poisson mean of expected lensed pairs during the observing period, for an expected lensing fraction $u$. Based on our simulations (Appendix~\ref{sec:BBH_simulations}), we assume that a fraction $u=1.5\times10^{-3}$ of all detectable events will be strongly lensed. This corresponds to a $20\%$ probability of having a lensed event in an O4-like catalog.

Figure~\ref{fig:ROC_cat_sig} shows both $\eta$ and detection probability against catalog significance level.  There is a 65\% chance that a lensed event, if present, can be correctly identified with PO2.0 with a significance $\geq 2.25\catsig$, a nearly two-fold improvement over \cite{haris2018identifying}. Combined with the intrinsic lensing probability, this leads to a $13\%$ chance that we can detect at least one lensed pair in O4 with a significance $\geq 2.25\catsig$ using our new statistic. On the other hand, a statistic such as \cite{haris2018identifying}'s $\bo\times \rlu$, which does not include informative population priors except for the time delay, has a 7.5\% chance of confidently detecting lensing. This demonstrates the importance of informative prior modeling and inclusion of information contained in all the inferred parameters of the binary for improving our chances of detection of strong lensing as highlighted in the literature \citep{cheung2023mitigating, janquart2023ordering, more2022improved}.

\section{Parameter estimation of lensed events}
\label{sec:po2_pe}

\begin{figure}[tbh]
\centering
\includegraphics[width=0.9\columnwidth]{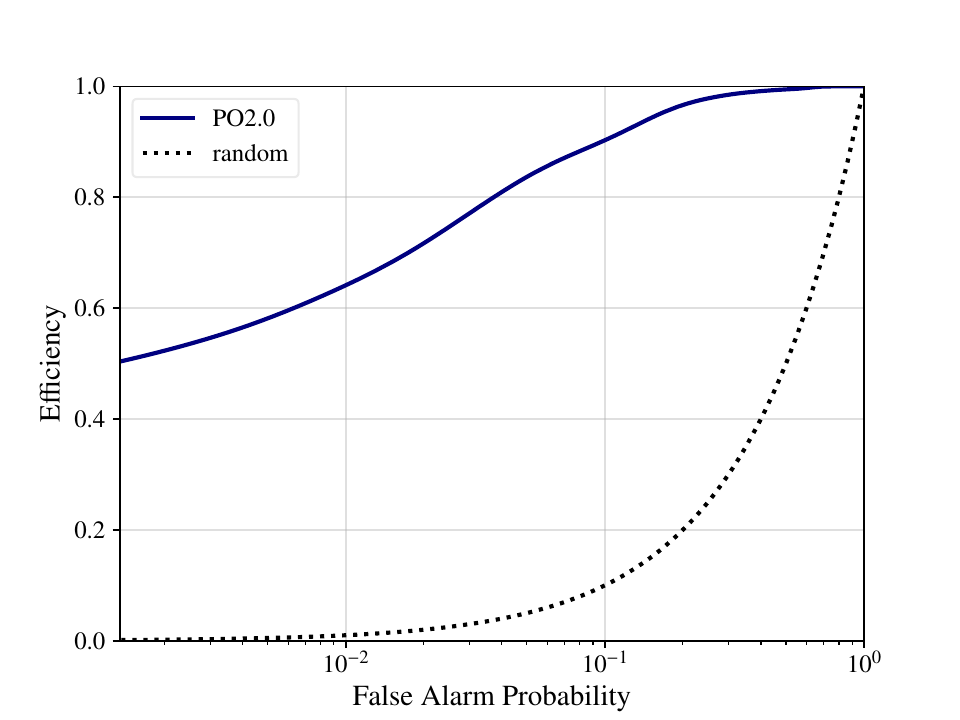}
\caption{ROC curves showing the efficiency of the PO2.0 statistic in identifying the correct Morse phase difference between two lensed images as a function of the false alarm probability. Here the efficiency (false alarm probability) is defined as the probability of the Bayes factor ${\blu_{n=0}}/{\blu_{n=1}}$ preferring the right (wrong) Morse phase among a set of lensed events.}
\label{fig:ROC_Morse_phase}
\end{figure}

In the previous section, we have demonstrated the power of PO2.0 to correctly identify strongly lensed events. In this section, we show how PO2.0 may also be used to estimate the combined posterior probability distributions of the lensed event's parameters, just like the (computationally expensive) joint PE methods.

We assume that a strong lensing search has already been performed using the PO2.0 statistic and that we have identified a pair of GW signals that are very likely to be lensed images of the same BBH merger. The posteriors on the signal and lens parameters are then given by the integrand of the numerator of eq.~\ref{eq:BLU_expression}. That  is,
\begin{eqnarray}
\label{eq:posterior_exp}
P(\theq, & \thbI,& \dthb \mid d_1, d_2, \HL) \propto  \dfrac{P(\theq, \thbI \mid d_1)}{P_\textsc{PE,1}(\theq, \thbI)} \\
&& \dfrac{P(\theq, \thbII=\thbI + \dthb \mid d_2)}{P_\textsc{PE,2}(\theq, \thbII=\thbI + \dthb)} ~~ P(\theq, \thbI, \dthb \mid \HL)  \nonumber.
\end{eqnarray}

We can now remove some of the simplifying approximations used in the calculation of $\blu$ (refer to Appendix~\ref{sec:BLU_simplification}), and focus more on accuracy to avoid potentially biasing the posteriors. We still work in the dominant mode approximation to the GW waveform, and we consider only two Morse phase differences, 0 and $\pi/2$.

\subsection{Morse phase difference}

We treat the estimation of Morse phase as a model selection problem between the $\dphi=0$ ($n=0$) and $\dphi=\pi/2$ ($n=1$) hypotheses. The corresponding Bayes factor is the ratio of the lensed evidences computed assuming Morse indices of $n = 0$ and $n = 1$. This is simply given by ${\blu_{n=0}}/{\blu_{n=1}}$, where $\blu_{n}$ is the lensing Bayes factor computed assuming a Morse index of $n$ (see eq.~\ref{eq:BLU_n_definition}). Figure~\ref{fig:ROC_Morse_phase} shows the ROC curves for identifying the Morse phase of a lensed event using this Bayes factor. Evidently, PO2.0 can correctly identify the Morse phase difference between the two images for $\gtrsim 80\%$ ($50\%$) of the lensed events with $0.05$ ($0.003$) false alarm probability (equivalent to $2\sigma $ ($3\sigma$) confidence).  We can also compute the posterior probability of each $\dphi$
\begin{multline}
\label{eq:Morse_phase_posterior}
P(\dphi=n{\pi}/{2} \mid d_1,d_2,\HL) \propto {\blu_n ~ P(\dphi=n{\pi}/{2} \mid \HL)},
\end{multline}
where $ P(\dphi=n{\pi}/{2} \mid \HL)$ is given by eq.~\ref{eq:dphi_prior}.

\begin{figure}[t]
\centering
\includegraphics[width=\columnwidth]{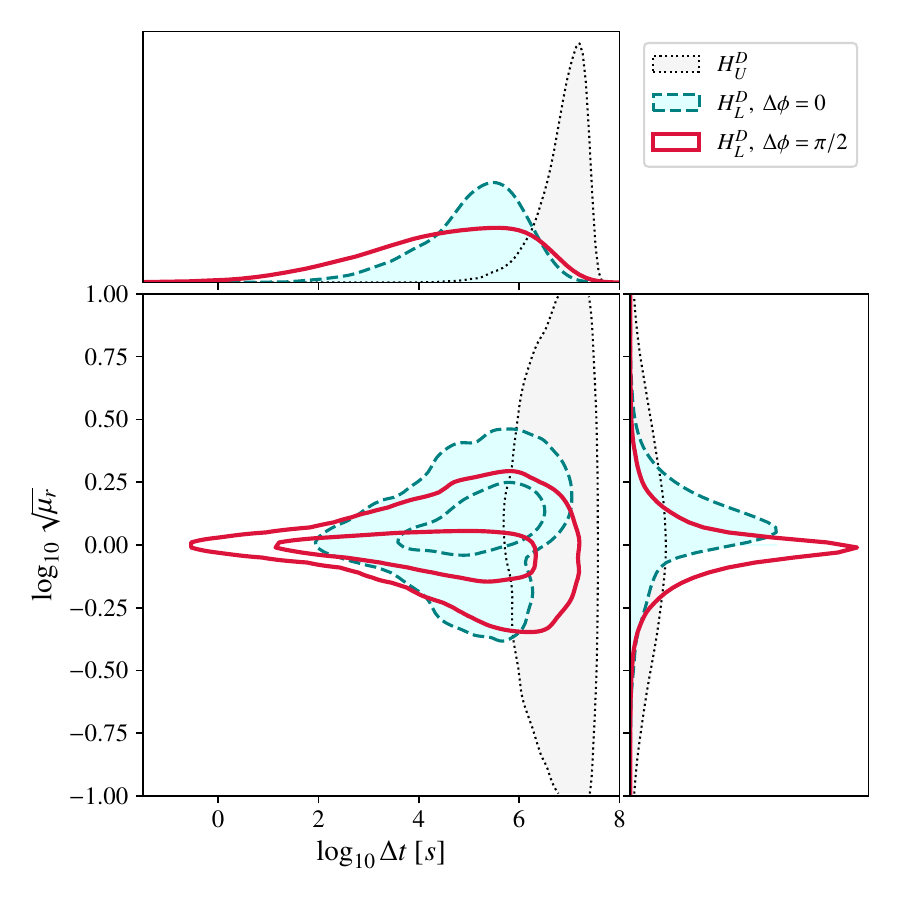}
\caption{The astrophysical prior $P(\log_{10}{\dt}, \log_{10}{\sqrt{\mur}} \mid \mathcal{H})$ on the relative magnification and time delay between pairs of detectable unlensed ($\mathcal{H} = \HU^D$) events, and between detectable lensed ($\mathcal{H} = \HL^D$) images. The top and side panels show their marginalized distributions. These distributions are marginalized over all other parameters.}
\label{fig:det_population_mur_dt_Morse}
\end{figure}

\subsection{Source parameters}

Assuming delta function posteriors for $t_{c,1}$ and $t_{c,2}$, we can marginalise the posterior given in eq.~\ref{eq:posterior_exp} over $t_{c,1}$ and $\Delta t$ (see Appendix~\ref{se:RLU_factorisation}). Also, as argued in Appendix~\ref{sec:phic_deltaphi_factorisation}, we can ignore the correlation of $\phi_c$ and $\psi$ with other parameters, and marginalise the posteriors over $\phi_c$ and $\psi$, resulting in
\begin{eqnarray}
P(\theq, d_{L,1}, && \mur  | d_1, d_2, \HL)  \propto  \sum_{n = 0,1} P(\Delta \phi = n\pi/2 | \HL) \,  \rlu_n \, \plu_n \,  \nonumber \\
 &&  P(\theq, d_{L,1}, \mur \mid \dt^\rom, \dphi=n\pi/2, d_1, d_2, \HL),
\end{eqnarray}
where $\rlu_n$ and $\plu_n$ are given by eqs.~\ref{eq:RLU_raw_prior_ratio} and \ref{eq:PLU_definition}. The (marginalized) prior $P(\mur, \Delta t \mid \dphi=n\pi/2, \HL)$ is shown in Fig.~\ref{fig:det_population_mur_dt_Morse}. Note that now $\theq$ vector does not include the polarisation angle $\psi$; still, we use the same symbol  $\theq$ for simplicity of notation. Above,
\begin{eqnarray}
P(\theq, d_{L,1},  \mur  & | &  \dt^\rom, \dphi=n\pi/2,  d_1, d_2, \HL) = \dfrac{P(\theq, d_{L,1}\mid d_1)}{P_\textsc{PE,1}(\theq,  d_{L,1})} \nonumber \\
& & \dfrac{P(\theq, d_{L,1} / \sqrt{\mur} \mid d_2)}{P_\textsc{PE,2}(\theq, d_{L,1} / \sqrt{\mur})} \nonumber \\
& & P(\theq,  d_{L,1}, \mur \mid \dt^\rom, \dphi=n\pi/2, \HL).
\end{eqnarray}
This distribution can be sampled by drawing samples of $\theq, d_{L,1}$ from $P(\theq, d_{L,1} \mid d_1)$, along with samples of $\mur$ from $Q(\mur)$, and then reweighting these samples by
\begin{eqnarray}
\label{eq:importance_weights}
\dfrac{1}{P_\textsc{PE,1}(\theq, d_{L,1})}~&& ~ \dfrac{P(\theq,  d_{L,1} / \sqrt{\mur} \mid d_2)}{P_\textsc{PE,2}(\theq, d_{L,1} / \sqrt{\mur})} \\
&& \frac{P(\theq, d_{L,1}, \mur \mid  \dt^\rom, \dphi=n\pi/2, \HL)}{Q(\mur)} . \nonumber
\end{eqnarray}
It is easy to see that this is formally independent of the choice of $Q(\mur)$. Note that this is the method that we followed (Appendix~\ref{sec:BLU_evaluation}), to calculate the evidence of $\HL$ through the Monte Carlo average of the samples described above.  If required, we can obtain the lensed posterior and evidence in one go.
\begin{figure*}[t]
\centering
\includegraphics[width=1.9\columnwidth]{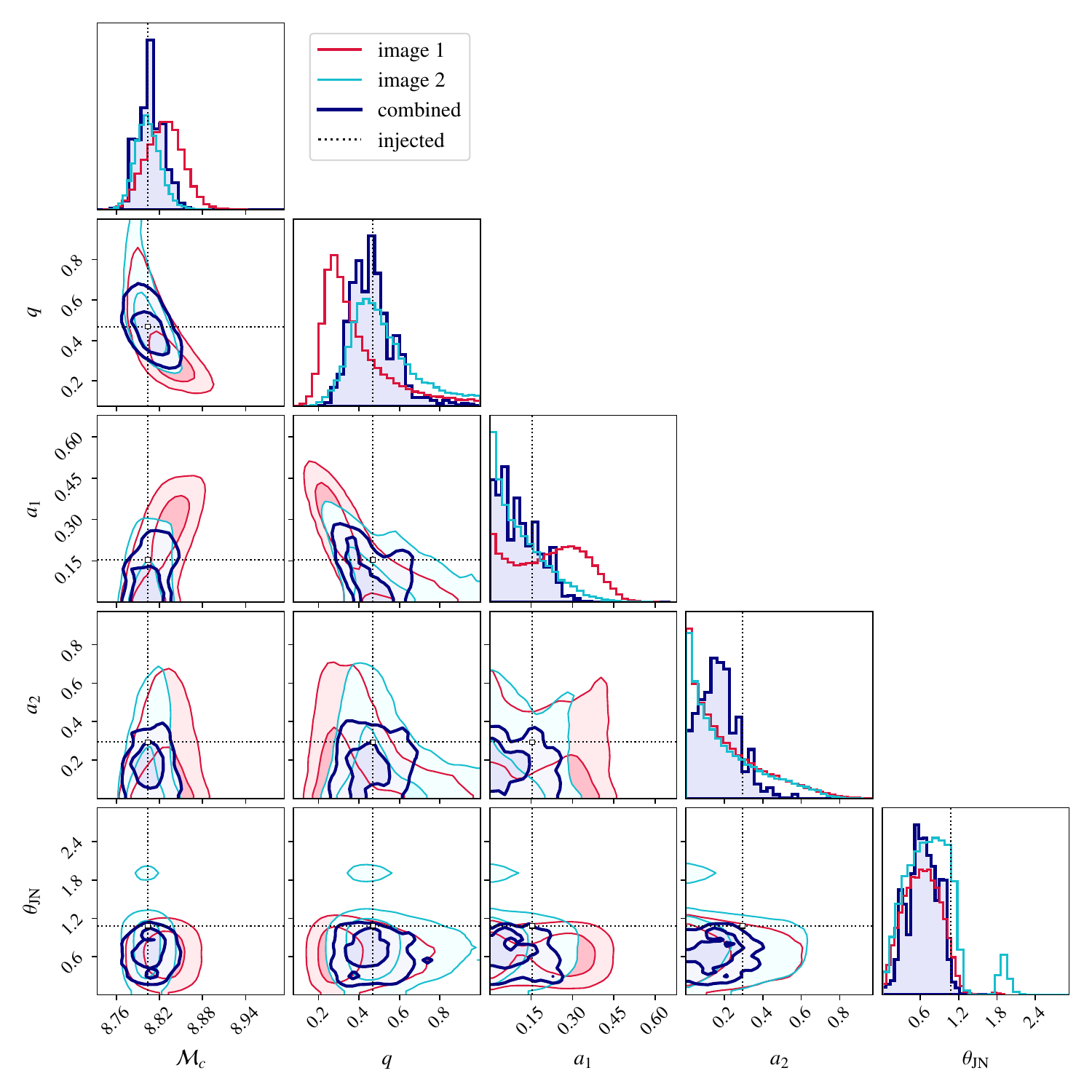}
\caption{Posterior probability distribution of the intrinsic parameters (chirp mass $\mathcal{M}_c$, mass ratio $q$, spin magnitudes $a_1$, $a_2$) and inclination $\theta_\mathrm{JN}$. These are shown for the two images of a simulated lensed event, along with those obtained by combining them with appropriate weights. The contours correspond to the 50 and 90 \% credible regions.}
\label{fig:PO2_PE_intrinsic_corner}
\end{figure*}
In Fig.~\ref{fig:PO2_PE_intrinsic_corner}, we show that the combined posteriors on the intrinsic parameters and inclination are tighter than those of the individual images. This shows that PO2.0 can be used to obtain tighter constraints on the lensed events' parameters at low computational cost, done simply by reweighting the first image's posterior.

\subsection{Sky localization}
\begin{figure}[t]
\centering
\includegraphics[width=\columnwidth,trim={0 1cm 0 1cm},clip]{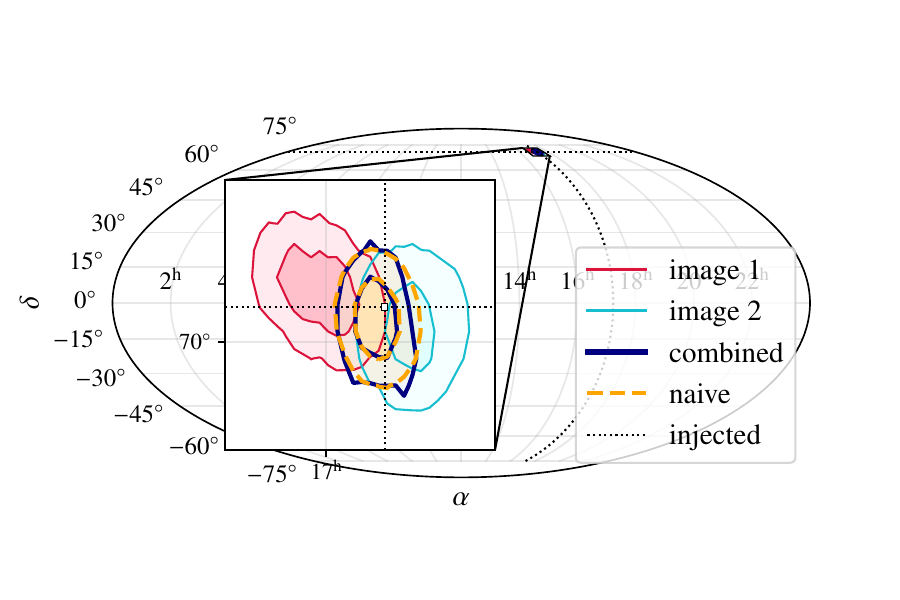}
\caption{Posterior probability distribution of the sky localization for the two images of a simulated lensed event, along with that obtained by combining them with appropriate weights. The contours correspond to the 50 and 90 \% credible regions. The posterior obtained by naively multiplying the two individual sky likelihoods is also shown.}
\label{fig:PO2_PE_sky_Mollweide_withNaive}
\end{figure}

Equation~\ref{eq:importance_weights} provides the posterior weights of all the parameters of the lensed event, including the sky location. However, one may note that the individual sky location posteriors and priors are largely uncorrelated with the rest of the parameters (barring the arrival time and time delay, but these are measured far too precisely for their priors to be non-trivial within their region of uncertainty). Statistically, then, their relative posterior weights are also uncorrelated with those for the rest of the parameters. Therefore, the combined sky location posterior is quite well approximated by just the product of its individual posteriors, as one would naively do assuming the two to be independent. We show in Fig.~\ref{fig:PO2_PE_sky_Mollweide_withNaive} that this is indeed the case. This conclusion may be useful when searching for electromagnetic counterparts to strongly lensed GW events at low latency.

\subsection{Relative magnification}

\begin{figure}[t]
\centering
\includegraphics[width=\columnwidth]{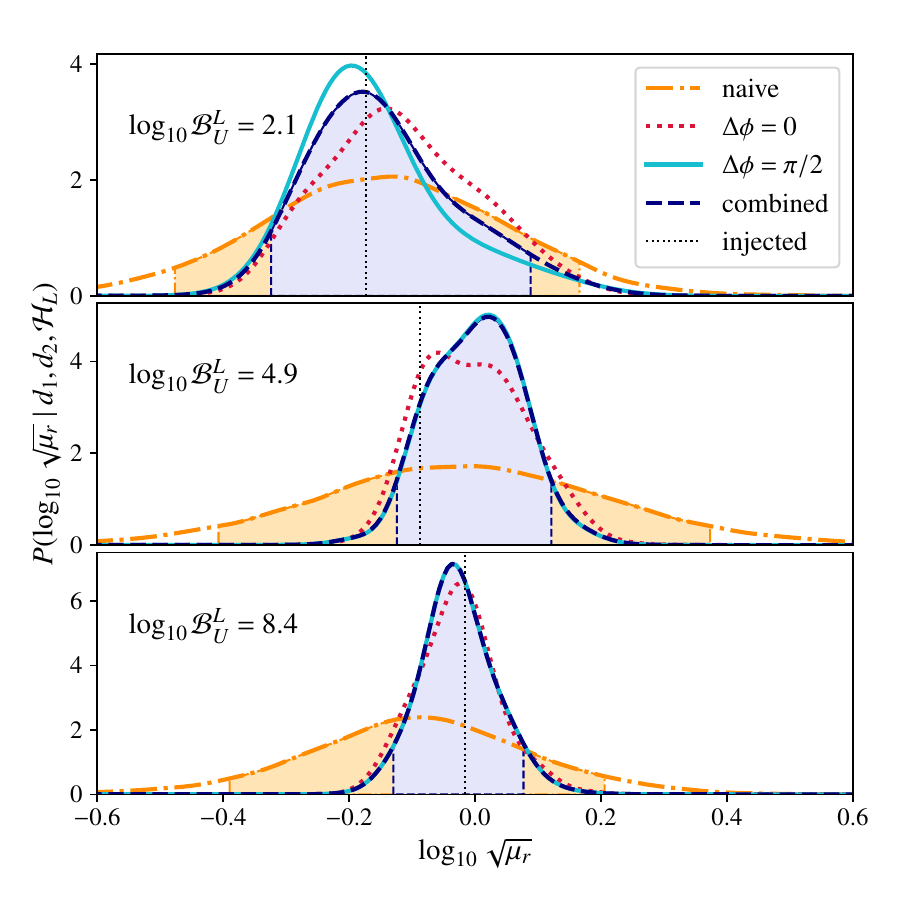}
\caption{Posterior probability distribution of the relative magnification between simulated pairs of lensed images. The three panels correspond to three different pairs of lensed events (the Bayes factor for each pair is indicated). Posteriors obtained by naively calculating the ratio distribution of the two events' $d_L$ posteriors, along with those calculated using the PO2.0 formalism under $\dphi=0$ and $\dphi=\pi/2$ hypotheses, and the posterior obtained by combining them with appropriate weights are shown. For the first and last of these, 90\% credible regions are also shown in shaded regions.}
\label{fig:log10_sqrt_mur_posteriors}
\end{figure}

Figure~\ref{fig:log10_sqrt_mur_posteriors} shows the posterior distributions of the relative magnification $\mur$ for three different lensed events (pairs of images) having small, moderate, and high lensing Bayes factors $\blu$. In this case, the combined posterior is better constrained than what one would naively obtain by finding the ratio distribution of the two signals' $d_L$ posteriors while treating them as independent probability distributions. This additional constraining power comes entirely because of the informative population prior, since without it, the weights in eq.~\ref{eq:importance_weights} are essentially just the product of likelihoods. The magnification posteriors tend to get narrower with $\blu$, which is due to the fact that these posteriors get narrower with higher S/N of the events, and because the $\blu$ itself increases with S/N.

\begin{figure}[t]
\centering
\includegraphics[width=\columnwidth]{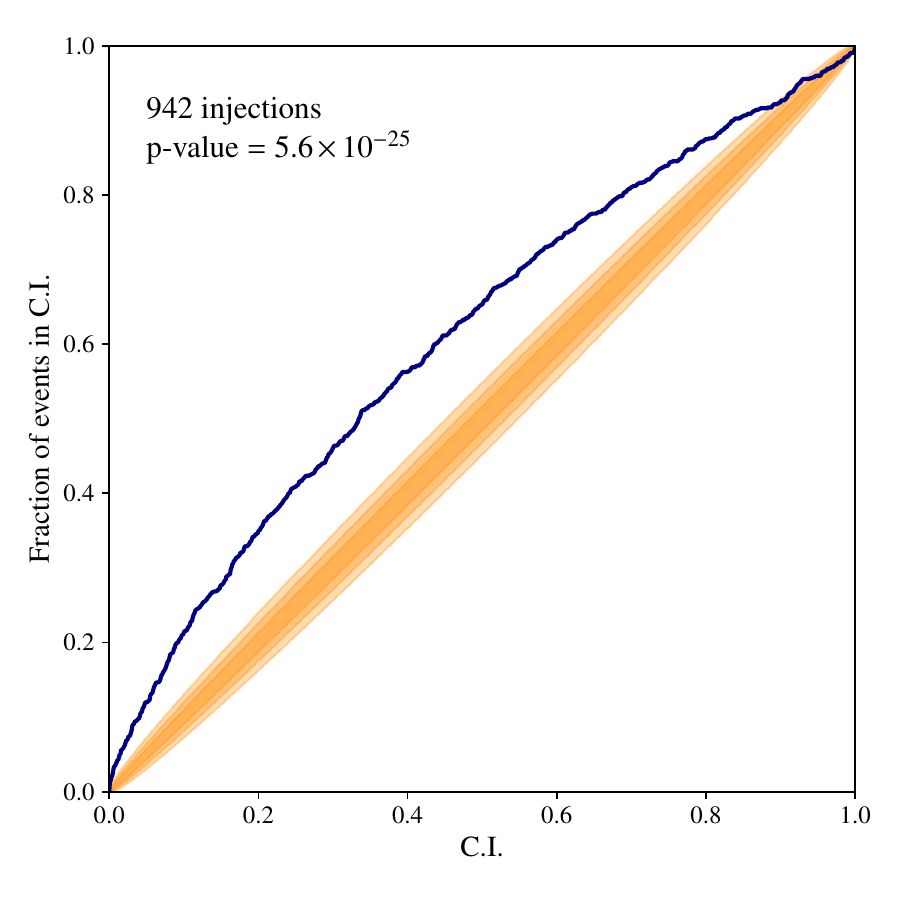}
\caption{$p-p$ plot showing the fraction $p\%$ of the 942 injections whose true value for the relative magnification lies within $p\%$ credible interval of the estimated posteriors. Orange bands show the 1, 2, and 3 $\sigma$ credible regions according to the p-value computed using the KS test.}
\label{fig:PP_PO2_PE_mur}
\end{figure}

In Fig.~\ref{fig:PP_PO2_PE_mur}, we draw a $p-p$ plot (see Appendix~\ref{sec:BBH_simulations} for further explanation) to show the performance of our PE scheme in estimating $\mur$. We find that $\mur$ is systematically overestimated, though the bias is small ($\sim 10\%$)\footnote{this was estimated by finding the amount of shift required in each of the $\log_{10}{\sqrt{\mur}}$ posteriors to make the $p-p$ plot fall on the diagonal band, leading to a high p-value in the KS test.}, which is an order of magnitude smaller than the typical width of the posteriors.

We believe that these biases are shortcomings of using reconstructed density estimations in our formalism. Since the combined posterior consists of overlapping regions in the two posteriors, only a small fraction of samples in the first signal's posterior have significant weight in the combined posterior, which makes resampling noisy. Joint PE methods alleviate this problem by generating new samples in the overlapping region to achieve high-fidelity posteriors. We therefore conclude that, though the current PO2.0 formalism is capable of measuring relative magnifications, it is limited by numerical inaccuracy and one should prefer joint PE for this task.

\section{Conclusion}
\label{sec:conclusion}

Strongly lensed GW signals, in the geometric optics approximation, are identical to each other except for an overall magnification, time delay, and a phase offset. Identifying such events requires checking the similarity between all possible pairs of detected GW signals to find pairs that stand out of the background of unrelated (unlensed) pairs. In this paper, we have presented an improved Bayes factor statistic called PO2.0 that efficiently identifies lensed events based on the overlap between their individual posteriors.

PO2.0 accounts for all the available information in the posterior distributions, as well as the population of GW sources and selection effects, with the only assumption that the signals are non-overlapping and their noise is uncorrelated. It reduces to various other statistics present in the literature under appropriate approximations. In particular, the expression of our Bayes factor is mathematically equivalent to that of joint PE methods that analyze two signals together. The difference is that instead of using stochastic sampling to sample the joint likelihood, we evaluate it by post-processing the KDEs of individual posterior distributions. We expect our statistic to have efficiencies comparable to the joint PE while being computationally much cheaper to evaluate. This will enable us to estimate the significance of candidate (lensed) pairs by doing ``deep'' background simulations, which is very difficult to do for joint PE pipelines.

Through a deep background analysis, we showed that PO2.0 outperforms existing low-latency search/veto pipelines in terms of detection efficiency. The additional efficiency over the earlier posterior overlap method comes both from using the correct population priors with selection effects and from using information stored in the Morse phase, magnification, and time delay. At the lowest pairwise false alarm probability that we can currently probe, $\sim 2\times 10^{-6}$, PO2.0 can correctly identify 65\% of all lensed events.

By appropriately combining the posteriors from two lensed signals, PO2.0 also provides a computationally inexpensive means for estimating the source and lens parameters. We show that PO2.0 can measure the relative magnification better than what one would estimate assuming the two posteriors to be independent, though, as of now, it is curbed by numerical errors that overestimate the magnification by $\sim 10\%$. More interestingly, we find that it is possible to identify the correct Morse phase difference in 50\% of the lensed events at $> 3\sigma$ confidence.

Using astrophysical simulations of strong lensing of GWs, we forecast that there is a 20\% probability of finding at least one lensed event in 18 months of observation at O4 sensitivity. The probability of confidently \textit{identifying} that lensed event using our method is 65\% lower. Therefore, the overall chance of a marginally confident detection of strong lensing using GW data alone is 13\%. If such an image pair is found, we can currently only put a lower bound of 2.25$\sigma$ on its significance level due to our deep, yet limited background. We wish to probe even deeper backgrounds in the future to enable a confident detection at $3\sigma$ or $5\sigma$.

Although we anticipate that PO2.0 will be as efficient as joint PE in identifying strongly lensed GW events, this needs to be demonstrated.  This is a computationally expensive task, due to the cost of running a background analysis for joint PE. Such a study is currently underway. Our immediate goal for the future is to compare PO2.0 with joint PE to find how much efficiency we lose due to posterior reconstruction errors that our method suffers from. Apart from deep background studies with joint PE, and further tests for checking the accuracy of PO2.0 (for example, waveform systematics \citep{garron2023waveform}), various other future directions are also open. Since the lensed evidence (eq.~\ref{eq:BLU_numerator}) is sensitive to the underlying lens model, PO2.0 can potentially be used to distinguish between different lens models in a way akin to \cite{janquart2023ordering, seo2024inferring, wright2023determination} and \cite{poon2024galaxy}. We would also like to include galaxy clusters in our lens population modeling in the future.

The flip side of being sensitive to model assumptions is that PO2.0 depends on the population models of the BBH mergers, which, as of now, are not directly measurable at high redshifts. Therefore, in the future, it may be useful to marginalize the PO2.0 statistic over various astrophysical models of the merger rate density.

Our simulations show that there is a 44\% chance that there will be a third detectable image present in the data if one has already detected a pair. Note that our expression of the lensing Bayes factor (eq.~\ref{eq:BLU_expression}) can easily be generalized to include more than two images. Indeed, the estimation of background is then computationally more expensive since the background grows exponentially with the number of image tuples one wishes to check the consistency for. However, triplets and quadruplets of lensed images may be far less prone to false alarms \citep{ccalicskan2023lensing}. It is therefore high time that strong lensing search pipelines rigorously take these additional images into account, and we are working towards that goal with PO2.0.

\section*{Acknowledgments}
We are grateful to K. Haris for his careful review of this manuscript and useful comments. We also thank Tejaswi Venumadhav and Javier Roulet for guidance in running \texttt{cogwheel} and other discussions, Rico Lo and Jose Ezquiaga for their help in running \texttt{phazap}, and Otto Hannuksela for helpful discussions regarding false alarms, $\rlu$ and selection effects. We also thank members of the ICTS Astrophysical Relativity Group and LIGO-Virgo-KAGRA Lensing Working Group for useful discussions. Our research is supported by the Department of Atomic Energy, Government of India, under Project No. RTI4001. The numerical calculations reported in the paper were performed on the Alice computing cluster at ICTS-TIFR.

\appendix

\section{Derivation of the new posterior overlap $\blu$}
\label{sec:BLU_derivation}
In this Appendix, we provide a detailed derivation of eq.~\ref{eq:BLU_expression} starting from the likelihood ratio
\begin{equation}
\label{eq:BLU_definition}
\blu = \dfrac{P(d_1,d_2\mid \HL)}{P(d_1,d_2\mid \HU)}.
\end{equation}

\subsection{The denominator}
The ``unlensed'' evidence $P(d_1,d_2\mid \HU)$ is easy to simplify because under the hypothesis that the two signals are not lensed, $d_1$ and $d_2$ are independent and their joint likelihood is given by the product of individual likelihoods
\begin{equation}
\label{eq:BLU_denominator}
P(d_1,d_2\mid \HU) = P(d_1 \mid \HU) ~ P(d_2 \mid \HU).
\end{equation}
Above, $P(d_j \mid \HU),\ j\in\{1,2\}$ are the marginalized likelihoods of $d_j$ under the hypothesis that the signals come from an unlensed population (or, the Bayesian evidence of  $\HU$):
\begin{equation}
\label{eq:unlens_evid}
P(d_j \mid \HU) = \int d\theta\ P(d_j \mid \theta) ~ P(\theta \mid \HU).
\end{equation}

Using the Bayes theorem, we can write
\begin{equation}
\label{eq:Bayes_theorem_PE}
P(d_j \mid \theta) = P_\pe(d_j) ~ \dfrac{P(\theta \mid d_j) ~}{P_{\pe,j}(\theta)} ,
\end{equation}
where $P_\pe(d_j)$ is the evidence obtained from the individual signals using a simple prior $P_{\pe,j}(\theta)$
\begin{equation}
P_\pe(d_j) = \int d\theta\ P(d_j \mid \theta) ~ P_{\pe,j}(\theta).
\end{equation}
Plugging in eq.~\ref{eq:Bayes_theorem_PE} in eq.~\ref{eq:unlens_evid}, we have
\begin{equation}
\label{eq:unlensed_evidence}
P(d_j \mid \HU) = P_\pe(d_j)\ \int d\theta\ P(\theta \mid d_j) ~ \dfrac{P(\theta \mid \HU)}{P_{\pe,j}(\theta)}.
\end{equation}
Essentially, these are just the evidences of the individual data segments obtained using standard PE techniques but reweighted to the appropriate population of unlensed binaries.

\subsection{The numerator}
By splitting the parameters $\theta$ into $\{\theq, \thbI, \dthb\}$ (see the discussion below eq.~\ref{eq:lensd_wave_h2}), $P(d_1,d_2\mid \HL)$ can be expressed as a joint likelihood of $d_1$ and $d_2$ marginalized over all the independent parameters
\begin{multline}
\label{eq:BLU_numerator}
P(d_1,d_2\mid \HL) = \int d\theq\ d\thbI\ d\dthb ~~ P(d_1,d_2 \mid \theq, \thbI, \dthb) \\  P(\theq, \thbI, \dthb \mid \HL).
\end{multline}

Note that the joint likelihood is only a function of the data, the model, and noise properties. Given the values of $\theq,\ \thbI$ and $\dthb$, it does not explicitly depend on the population they originated from, and hence we have dropped $\HL$ from its arguments.

It is instructive to imagine the process that results in these two data $d_1$ and $d_2$. A single binary merges and the emitted GWs are strongly lensed, causing two nearly identical copies of the same GW waveform to appear at our detectors. The first of these copies has a waveform parameterized by $\theta = \{\theq, \thbI\}$, that gets added to a realization of random noise $n_1$, leading to a strain $d_1$. The second copy, on the other hand, has a waveform parameterized by $\theta = \{\theq, \thbII = \thbI + \dthb\}$, that gets added to a \textit{different} realization of random noise $n_2$, leading to a strain $d_2$. Now, if the two data $d_1$ and $d_2$ have no temporal overlap, and provided the noise realizations $n_1$ and $n_2$ are uncorrelated, we can factorize their joint likelihood into individual likelihoods
\begin{multline}
\label{eq:likelihood_factorization}
P(d_1, d_2\mid \theq, \thbI, \dthb) = P(d_1 \mid \theq, \thbI)  \\ P(d_2 \mid \theq, \thbII=\thbI + \dthb)
\end{multline}

Using the Bayes theorem (eq.~\ref{eq:Bayes_theorem_PE}), we can expand these individual likelihoods in terms of the posterior distribution and evidence obtained using a simple prior from standard PE analysis
\begin{align*}
P(d_1 \mid \theq, \thbI) &= P_{\pe}(d_1) ~ \dfrac{P(\theq, \thbI \mid d_1)}{P_{\pe,1}(\theq, \thbI)},&\\
P(d_2 \mid \theq, \thbII = \thbI + \dthb) &= P_{\pe}(d_2) ~ \dfrac{P(\theq, \thbI + \dthb \mid d_2)}{P_{\pe,2}(\theq, \thbI + \dthb)} &.
\end{align*}
Substituting these expressions in eq.~\ref{eq:BLU_numerator}, we get
\begin{multline}
\label{eq:lensed_evidence}
P(d_1,d_2\mid \HL) = P_{\pe}(d_1)  \, P_{\pe}(d_2)\  \\
\int d\theq\ d\thbI\ d\dthb ~ P(\theq, \thbI, \dthb \mid \HL) \\
\left[ \dfrac{P(\theq, \thbI \mid d_1)}{P_{\pe,1}(\theq, \thbI)}\ \dfrac{P(\theq, \thbI + \dthb \mid d_2)}{P_{\pe,2}(\theq, \thbI + \dthb)}  \right]
\end{multline}
and finally, substituting eqns.~\ref{eq:unlensed_evidence} and \ref{eq:lensed_evidence} in eq.~\ref{eq:BLU_definition}, we obtain the strong lensing Bayes factor that incorporates informed priors regarding binary populations and includes all the available information about the binary parameters. This is given in eq.~\ref{eq:BLU_expression}.

\section{Further simplifications to $\blu$}
\label{sec:BLU_simplification}

To confidently identify strongly lensed events, it is important not only to adopt a good detection statistic but also to evaluate it accurately and precisely. Numerical noise in the calculation of the statistic can ``smear'' its distribution under the unlensed and lensed hypotheses, increasing false alarm probability and decreasing efficiency. It is therefore important that we analytically simplify our expression for the Bayes factor in eq.~\ref{eq:BLU_expression} in order to minimize statistical noise to achieve higher efficiency. In this appendix, we show how this can be achieved by ignoring the PE priors and factorizing the rest into chunks of lower-dimensional integrals that are less noisy and easier to evaluate. In this process, we will also show how integrals over arrival time and time delay naturally factorize into $\rlu$ from \cite{haris2018identifying}.

\subsection{PE priors}
PE of GW signals is computationally expensive. Therefore, the assumed joint prior on BBH parameters is often chosen to be a product of independent priors that are easy to evaluate. The most common choice is that of having uniform priors for all parameters except the luminosity distance, which is chosen to be uniform in comoving volume instead.

\begin{figure}[t]
\centering
\includegraphics[width=\columnwidth]{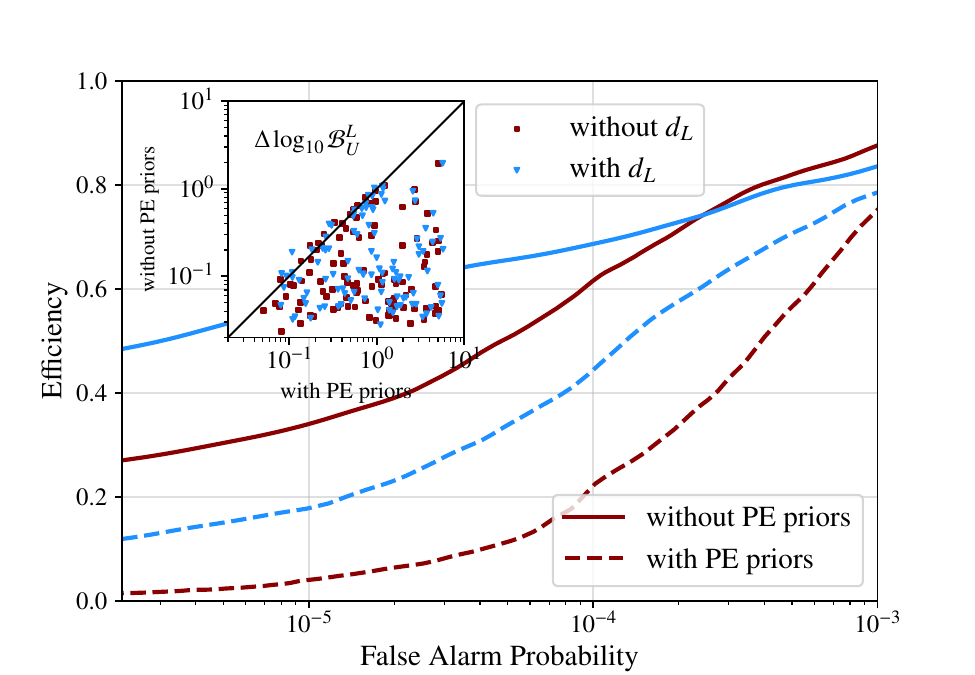}
\caption{ROC curves showing improvement in detection efficiency due to ignoring PE priors in eq.~\ref{eq:BLU_expression}. The inset shows bootstrap errors (estimated by repeatedly resampling from the available samples and computing the Bayes factor) in the calculation of individual $\blu$'s with and without PE priors. These $\blu$'s were calculated assuming population priors including selection effects ($H_U^D$ and $H_L^D$).}
\label{fig:ROC_noPE}
\end{figure}

If the PE prior is indeed uniform, it simply drops out of eq.~\ref{eq:BLU_expression} \citep{cheung2023mitigating}. While this is not strictly true if the luminosity distance is also integrated over, we find that, empirically, there is not much to gain from keeping the correct prior inside the integrals. Figure~\ref{fig:ROC_noPE} shows that, in fact, keeping the (non-uniform) PE priors makes the calculations much more noisy, leading to a loss in sensitivity. We therefore drop the PE prior for all the BBH parameters, including luminosity distance.

\subsection{Sky location}
BBHs are expected to be distributed isotropically in the sky, and a 3-detector HLV network is almost equally sensitive to all parts of the sky when averaged over a length of observing time sufficiently longer than a day. This implies that the population prior of sky localization parameters is isotropic in the sky, with little distortion due to selection effects. Furthermore, the posterior distributions of right-ascension ($\alpha$) and declination ($\delta$) are mostly uncorrelated with those of the remaining BBH parameters. These two facts lead to the conclusion that we can factorize out the integrals over sky parameters into a simple sky overlap $\slu$, obtained by putting $\theq = \{\alpha, \sin{\delta}\}$ and $P(\theq) = \mathrm{Uniform}(\alpha, \sin{\delta})$ in eq.~\ref{eq:Bo_haris_et_al}
\begin{equation}
\label{eq:skyoverlap}
\slu = 4\pi \int\limits_0^{2\pi} d\alpha \int\limits_{-1}^{1} d\sin{\delta}\ P(\alpha, \sin{\delta} \mid d_1) ~ P(\alpha, \sin{\delta} \mid d_2)
\end{equation}

\subsection{Coalescence phase and Morse phase}
\label{sec:phic_deltaphi_factorisation}

Next, we turn to the Morse phase $\dphi$, whose population prior distribution $P(\dphi={n\pi}/{2} \mid \HL)$ is discrete, where $n = \{0,1,2\}$. Clearly, this will not change the unlensed evidences, which have no dependence on the lensing prior. The lensed evidence, on the other hand, will now be a weighted sum over different Morse phases, with weights given by $P(\dphi={n\pi}/{2} \mid \HL)$.

The next simplification comes by noting that the population prior on $\phi_c$ and $\psi$ is uniform, and remains so even after considering selection effects. Furthermore, the posterior distributions of $\phi_c$ are mainly correlated with $\psi$, and not with the rest of the BBH parameters. While none of this holds if there are significant contributions from higher modes of GW radiation \citep{janquart2021identification, vijaykumar2023detection}, such cases are sufficiently rare that we can handle them individually. For the vast majority of signals, then, we can factorize the integrals over $\phi_{c,1}$ and $\psi$ in a similar fashion as we did for sky overlap by introducing $\plu_n$
\begin{equation}
\label{eq:PLU_definition}
\plu_n = 4\pi^2 \int\limits_0^\pi d\psi \int\limits_0^{2\pi} d\phi_{c_1} ~ P(\phi_{c,1}, \psi \mid d_1) ~ P(\phi_{c,1}+\dfrac{n\pi}{2}, \psi \mid d_2),
\end{equation}
and the Bayes factor becomes
\begin{equation}
\label{eq:Morse_phase_factorization}
\blu = \b  ~ \slu ~ \sum_n P(\dphi=\dfrac{n\pi}{2} \mid \HL) ~ \plu_n.
\end{equation}
where $\b$ is the Bayes factor computed without considering the parameters $\{\alpha, \sin\delta, \phi_{c,1}, \dphi, \psi\}$. Note that, in the final step, we have also ignored correlations between the Morse phase and magnification ratio by making $\b$ independent of $\dphi$. This approximation is empirically motivated at this point\footnote{We have a finite set of samples for the lensed prior, with the number of samples reducing even further if we divide this set into those having $\dphi=0$ and those having $\dphi=\pi/2$. This makes the integrals more noisy if the lensing prior is conditioned on $\dphi$.}, but it is easy to relax if so desired, to
\begin{equation}
\label{eq:Morse_phase_NOT_factorized}
\blu = \slu ~ \sum_n P(\dphi=\dfrac{n\pi}{2} \mid \HL) ~ \b_n ~ \plu_n,
\end{equation}
where $\b_n$ is evaluated with the lensed population prior conditioned on the Morse phase.

\subsection{Arrival time and time delay}
\label{se:RLU_factorisation}

Here we show how the arrival time and time delay may be integrated out from eq.~\ref{eq:BLU_expression} in the form of $\rlu$. In the calculations that follow, we shall keep the PE priors, sky parameters, coalescence phase, and polarization angle for the sake of generality, and factor them out again towards the end.

The measurement error (i.e. the width of the marginalized posterior distribution) in the arrival time is $\sim$ milliseconds. On the other hand, lensing time delays are usually of the order of minutes to months. Thus, one could approximate the arrival time posterior by a delta function at the measured value: $P(t_{c} \mid d_j) \simeq \delta(t_{c,j}-t_{c,j}^\rom)$. This implies that the time delay posterior will also be a delta function at the measured value of the time delay, $P(\dt \mid d_1, d_2) \simeq \delta(\dt-\dt^\rom) = \delta(\dt-t_{c,2}^\rom + t_{c,1}^\rom)$.

Writing $\thbI = \{\thbI', t_{c,1}\}$ and $\Delta \thb = \{\Delta \thb', \Delta t\}$, and substituting these delta function posteriors in the expression for the lensed evidence, eq.~\ref{eq:lensed_evidence} for a fixed Morse phase difference, $ \dphi = n \pi/2$, leads to
\begin{align}
P(d_1, & d_2 \mid \dphi = n \pi/2, \HL) = P_{\pe}(d_1) ~ P_{\pe}(d_2) \nonumber &\\
& \int d\theq\ d\thbI'\ d\mur\ dt_{c,1}\ d\dt \nonumber &\\
& \left[ \vphantom{\int_1^2} \right. \dfrac{P(\theq, \thbI' \mid d_1) ~ \delta(t_{c,1}-t_{c,1}^\rom)}{P_{\pe,1}(\theq, \thbI' \mid t_{c,1})~{P_{\pe,1}(t_{c,1})}} \nonumber &\\
&~ \dfrac{P(\theq, \thbI' + \dthb' \mid d_2) ~ \delta(t_{c,1}+\dt-t_{c,1}^\rom-\dt^\rom)}{P_{\pe,2}(\theq, \thbI' + \dthb' \mid t_{c,1}+\dt) ~ P_{\pe,2}(t_{c,1}+\dt)} \nonumber &\\
&~ P(\theq, \thbI', \mur \mid t_{c,1}, \dt, \dphi = n \pi/2, \HL) \nonumber &\\
& \left. P(t_{c,1}, \dt \mid \dphi = n \pi/2, \HL) \vphantom{\displaystyle\int} \right], &
\end{align}
where $\thbI'=\{d_{L,1}, \phi_{c,1}\}$ and $\dthb'=\{\mur, \dphi=n\pi/2\}$. Note that the lensed population prior has been split into the joint probability of $t_{c,1}, \dt, \dphi$ and a part conditioned on them.  This preserves the correlations between $dt$ and $\dphi$, i.e. it accounts for the fact that the distribution of lensed time delays is different for each Morse phase difference (see Fig.~\ref{fig:det_population_mur_dt_Morse}).

It is natural to assume that the priors on arrival time will be uncorrelated with all the other BBH parameters, resulting in
\begin{multline}
\label{eq:lensed_evidence_RLU}
P(d_1,d_2\mid \dphi=n\pi/2, \HL) = \dfrac{P(t_{c,1}^\rom, \dt^\rom \mid \dphi=n\pi/2, \HL)}{P_{\pe,1}(t_{c,1}^\rom)~ P_{\pe,2}(t_{c,1}^\rom+\dt^\rom)} ~ \\
P_{\pe}(d_1) ~ P_{\pe}(d_2) \int d\theq\ d\thbI'\ d\mur \\
\left[ \vphantom{\int_1^2} \dfrac{P(\theq, \thbI' \mid d_1)}{P_{\pe,1}(\theq, \thbI')} ~~ \dfrac{P(\theq, \thbI' + \dthb' \mid d_2)}{P_{\pe,2}(\theq, \thbI' + \dthb')} \right.\\
\left. ~ P(\theq, \thbI', \mur \mid \dt^\rom, \dphi=n\pi/2, \HL)  \vphantom{\int_1^2} \right].
\end{multline}
Notice that the integrand is almost identical to that of eq.~\ref{eq:lensed_evidence}. The only difference is that the population prior $P(\theq, \thbI', \mur~\mid~\dt^\rom, \dphi= n\pi/2, \HL)$ is now conditioned on the measured value of time delay, apart from the Morse phase difference of $n\pi/2$.

For the unlensed evidence of the first image, we start by splitting $\theta$ as $\{\theta', t_c\}$, and substitute the delta functions in eq.~\ref{eq:unlensed_evidence} to obtain
\begin{multline}
\label{eq:unlensed_evidence_RLU}
P(d_j \mid \HU) = \dfrac{P(t_{c,j}^\rom \mid \HU)}{P_{\pe,j}(t_{c,j}^\rom)} ~ P_{\pe}(d_j) \\
~  \int d\theta'\ P(\theta' \mid d_j)~ \dfrac{P(\theta' \mid \HU)}{P_{\pe,j}(\theta')},
\end{multline}
where $\theta'$ are all BBH parameters except the arrival time $t_c$.

Putting together equations \ref{eq:BLU_definition}, \ref{eq:lensed_evidence_RLU}, and \ref{eq:unlensed_evidence_RLU}, the Bayes factor $\blu_n$ for each Morse phase difference  can be written as
\begin{equation}
\label{eq:BLU_n_definition}
\blu_n = \rlu_n ~ \blu'_n,
\end{equation}
where the integrals over arrival time are encapsulated in an overall factor $\rlu_n$ that is the ratio of population priors evaluated at the measured arrival time and time delay, for a given Morse phase difference of $n\pi/2$,
\begin{equation}
\label{eq:RLU_raw_prior_ratio}
\rlu_n = \dfrac{P(t_{c,1}^\rom, \dt^\rom \mid \dphi=n\pi/2, \HL)}{P(t_{c,1}^\rom \mid \HU) ~ P(t_{c,2}^\rom \mid \HU)}
\end{equation}
and $\blu'_n$ is the Bayes factor calculated without considering arrival time, time delay, and Morse phase difference, but with the lensed population prior conditioned on the measured time delay and a Morse phase difference of $n\pi/2$.

If we factor out the sky parameters, polarization angle, and coalescence phase, and ignore correlations between $\mur$ and $\dphi$ as done in the previous sections, $\blu_n$ becomes
\begin{equation}
\label{eq:BLU_n_definition_factored}
\blu_n = \b' ~ \slu ~ \plu_n ~ \rlu_n.
\end{equation}
In the above, $\b'$ is the Bayes factor computed without considering the parameters $\{\alpha, \sin\delta, \phi_{c,1}, \dphi, \psi, t_{c,1}, \dt\}$, calculated with the lensing population prior conditioned on measured time delay. With this, the total strong lensing Bayes factor is given by
\begin{align}
\label{eq:BLU_RLU_Morse_phase}
\blu &= \sum_n P(\dphi={n\pi}/{2} \mid \HL) ~ \blu_n.
\end{align}

We now show that eq.~\ref{eq:RLU_raw_prior_ratio} is identical to \cite{haris2018identifying}'s expression for the $\rlu$ if we impose their prior assumptions. We ignore all dependence on the Morse phase and assume that the unlensed arrival times are uniformly distributed within the observing duration $T$, i.e. $P(t_{c,i}^\rom \mid \HU) = 1/T$. Noting that the condition $t_{c,1} < t_{c,2}$ rules out half of the prior area in $\{t_{c,1},t_{c,2}\}$ space, we get
\begin{equation}
P(t_{c,1}^\rom \mid \HU) ~ P(t_{c,2}^\rom \mid \HU) = \dfrac{2}{T^2}.
\end{equation}
The arrival time for the first of the lensed images is also assumed to be uniform, but only within $[0,T-\dt^\rom]$ since, if it arrived after $T-\dt^\rom$, the second image would not fall within $[0,T]$ for $\dt^\rom>0$. Hence,
\begin{equation}
P(t_{c,1}^\rom \mid \dt^\rom, \HL) = \dfrac{1}{T-\dt^\rom}.
\end{equation}

Putting these details together, we find that $\rlu$ can be written as
\begin{equation}
\label{eq:RLU_Haris}
\rlu = \dfrac{P(\dt^\rom \mid \HL)}{P(\dt^\rom \mid \HU, T)},
\end{equation}
where
\begin{equation}
P(\dt^\rom \mid \HU, T) = 2\dfrac{T - \dt^\rom}{T^2}
\end{equation}
is the probability distribution of time delay in observing duration $T$ if the two arrival times result from a Poisson process.

\section{Evaluating the new $\blu$}
\label{sec:BLU_evaluation}
Here we discuss how to evaluate the Bayes factor (eq.~\ref{eq:BLU_RLU_Morse_phase}) on a pair of posterior samples. Before starting the full calculation, we perform some checks to quickly veto signal pairs that are highly unlikely to be lensed. The idea is that if the marginalized distributions of even a single parameter among $\theq$ do not overlap, the overall joint distribution will also have no overlap and the Bayes factor will be zero. For each $\theq$, one can check whether the prior ranges, the marginalized 1D posterior ranges, and their histograms overlap, and halt the calculation if they don't.

In principle, the same technique may also work in $\mathcal{N}$-dimensional marginalized posteriors of all possible n-tuples of $\theq$, but histograms become noisy and unreliable when the number of dimensions $\mathcal{N}\geq3$, so we only check for $^\mathcal{N}C_2$ 2D histogram overlaps in addition to 1D. With these checks, we can quickly veto out about 40\% of unlensed pairs within a second. If a pair of posteriors passes these checks, we proceed with the full calculation as described below.

\subsection{Calculating $\rlu_n$}
We estimate the observed time delay by taking the absolute difference of the median values of each marginalized posterior distribution of $t_c$. We then compute the numerator of eq.~\ref{eq:RLU_raw_prior_ratio} using \texttt{scipy}'s \texttt{gaussian\_kde} over samples of lensing time delays. The latter are obtained for each Morse factor from our astrophysical simulation (Appendix~\ref{sec:BBH_simulations}). The denominator is assumed to be a Poisson distribution as in eq.~\ref{eq:RLU_Haris}, calculated analytically with an observing duration of 18 months.

\subsection{Calculating $\slu$}
We compute 2D histograms of the posterior samples of $\{\alpha, \sin\delta\}$ from the two signals, multiply them, take a quadrature sum, and multiply it by $4\pi$.

\subsection{Calculating $\plu_n$}
While this could also be performed with a quadrature sum as above, we find that the result is more accurate if we use a KDE of the posterior distributions, and use importance sampling for integration. For each value of the Morse index $(n = 0, 1, 2)$, we therefore compute the Gaussian KDE $K_{2,n}$ of the 2D posterior distribution of $\phi_c + n\pi/2$ and $\psi$ from the second signal. We then evaluate $K_{2,n}$ over $N_1$ posterior samples of $\{\phi_c,\psi\}$ obtained from the first signal, and take their average
\begin{equation}
\plu_n = \dfrac{4\pi^2}{N_1}\sum\limits_k^{N_1} K_{2,n}(\phi_c^k, \psi^k).
\end{equation}

\subsection{Calculating $\b'$}
The expression for $\b'$ is formally the same as eq.~\ref{eq:BLU_expression}, though with fewer parameters being integrated over. Specifically, these are luminosity distance $\thbI=\{d_L\}$, magnification ratio $\dthb=\{\mu_r\}$, and the set $\theq=\{m_1^z, m_2^z, a_1, a_2, \cos\theta_{JN}\}$ including detector frame masses, spin magnitudes, and inclination. All of these parameters are tightly correlated with each other in either the posterior, the population prior, or both. Though we do not include spin orientation angles or eccentricity in our tests, they can be trivially incorporated if desired.

Due to the high dimensionality, we must resort to importance sampling techniques for integration. For the unlensed evidences in the denominator, we build a Gaussian KDE $K_U$ of the unlensed population prior using the samples obtained from our astrophysical simulation (Appendix~\ref{sec:BBH_simulations}). Ignoring the PE prior, we then evaluate it over the $N_j$ posterior samples obtained from the $j$'th signal ($j = 1,2$) before taking an average
\begin{multline}
\int d\theq\ dd_{L,j}\ P(\theq, d_{L,j} \mid d_j) ~ P(\theq, d_{L,j} \mid \HU) \\\simeq \dfrac{1}{N_j}\sum\limits_k^{N_j} K_U(\theq^k, d_{L,j}^k).
\end{multline}
The numerator involves the integration of the following product of terms over $\theq, d_{L,1}$ and $\mur$
$$P(\theq, d_{L,1} \mid d_1) ~ P\left(\theq, \dfrac{d_{L,1}}{\sqrt{\mur}} \mid d_2\right) ~ P(\theq, d_{L,1}, \mur \mid \HL)$$
To use importance sampling, we need samples of $\theq,\ d_{L,1}$ and $\mur$. The posterior distribution of the first signal provides samples of $\theq, d_{L,1}$ while for $\mur$, one may draw samples from any normalized probability distribution $Q(\mur)$ and compensate for it by doing a Monte Carlo average of the following quantity instead
$$P\left(\theq, \dfrac{d_{L,1}}{\sqrt{\mur}} \mid d_2\right) ~ \dfrac{P(\theq, d_{L,1}, \mur \mid \HL)}{Q(\mu_r)}$$
We choose $Q(\mur)$ to be the lensed population itself (marginalized over all parameters except $\mur$).

We build three Gaussian KDEs: $K_2$ for the samples of the second posterior, $K_L$ for samples of the lensing population prior, and  $K_{L,\mu}$ for $Q(\mur)$. We then evaluate the following average over the  $\theq, d_{L,1}$ samples of the first posterior, and $\mur$ samples drawn from $Q(\mur)$
\begin{multline}
\label{eq:lensed_evidence_eval}
\int d\theq\ dd_{L,1}\ d\mur ~ P(\theq, d_{L,1} \mid d_1) ~ P(\theq, \dfrac{d_{L,1}}{\sqrt{\mur}} \mid d_2)\\
~  P(\theq, d_{L,1}, \mur \mid \HL) \\
\simeq \dfrac{1}{N_1} \sum\limits_k^{N_1} \left[ K_2\left(\theq^k, \dfrac{d_{L,1}^k}{\sqrt{\mur^k}}\right) \right.
\left. ~ \dfrac{K_L(\theq^k, d_{L,1}^k, \mur^k)}{K_{L,\mu}(\mur^k)} \right],
\end{multline}
where $N_1$ is the number of samples used in the evaluation.

\begin{figure}[t]
\centering
\includegraphics[width=\columnwidth]{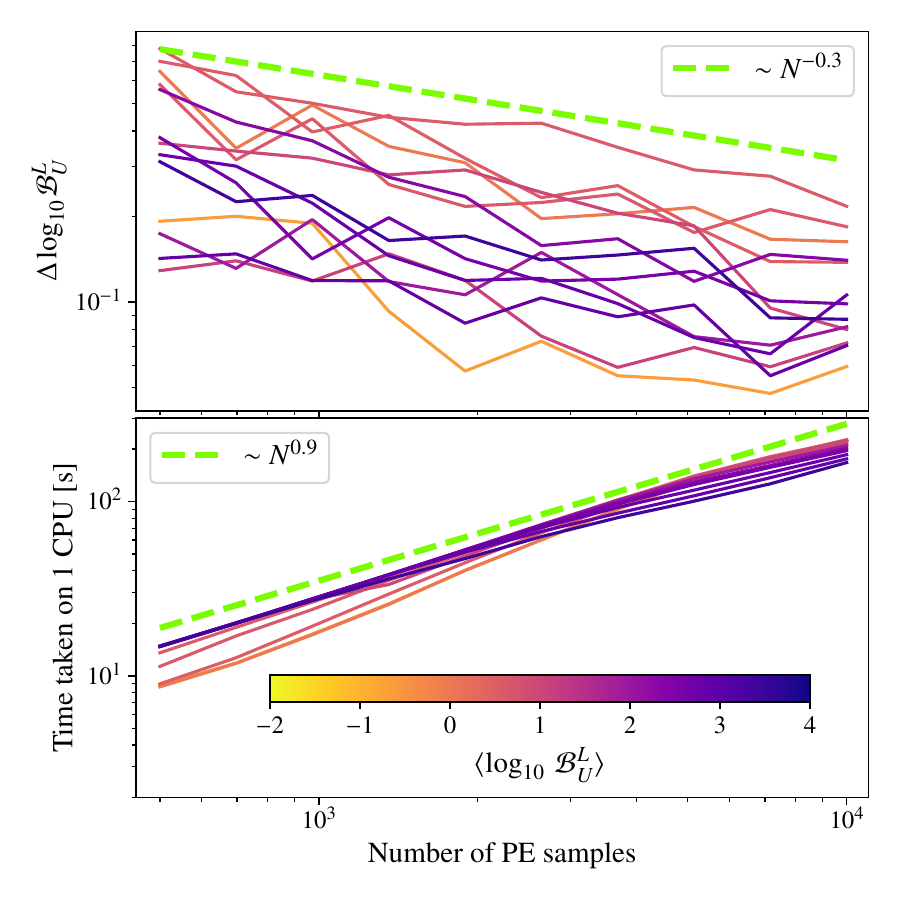}
\caption{The {upper} panel shows the bootstrap error (estimated by repeatedly resampling from the available samples, computing the Bayes factor, and taking the difference between their 16th and 84th percentile) in Bayes factor as we change the number of posterior samples used in the calculation, for a few different pairs of posteriors. For the same pairs, the {lower} panel shows the time taken to complete the calculation. The colorbar indicates the mean value of the Bayes factor for that pair of posteriors calculated by taking $10^4$ samples.}
\label{fig:Det_PO2_performance}
\end{figure}
The maximum dimensionality of the Gaussian KDE involved in these integrals is 7, which is small enough that we can reliably estimate the Bayes factor with $\sim 3 \times 10^4$ prior samples, and $\sim 10^4$ posterior samples. Figure~\ref{fig:Det_PO2_performance} shows that this method has a slow convergence resembling a power-law, and the time taken for each calculation increases sub-linearly with the number of posterior samples. The error is typically $\mathcal{O}(0.1)$ when one uses $\sim 10^4$ posterior samples. This is at the same order as the error one would encounter by estimating the lensed and unlensed evidences using Nested Sampling over the joint parameter space by assuming a typical tolerance on the evidence $Z$ of $d\log_e Z \lesssim 0.1$.

\section{Astrophysical simulations of BBH mergers}
\label{sec:BBH_simulations}

All the simulations in this paper were performed assuming a flat $\Lambda$CDM cosmology with parameters $\Omega_{\mathrm{m},0}=0.3$ and $h=0.7$.

\subsection{Unlensed BBHs}

We adopt a similar methodology as \cite{barsode2024constraints} to simulate a population of unlensed GW sources. We focus on BBH mergers since they are currently the dominant sources observable by the LIGO-Virgo detectors, and are also more likely to be lensed due to their larger horizon for detection \citep{wierda2021beyond}. This requires simulating the BBH population to a redshift of 7 in order to capture most of the lensed events at the current sensitivity of GW detectors. Since the merger rate density at redshifts $\gtrsim 1.5$ is largely unconstrained by observations \citep{abbott2023population}, we adopt the \cite{madau2014cosmic} model that assumes that BBHs follow the star formation rate.

We randomly draw $10^8$ BBH mergers with masses distributed according to the ``power law + peak'' model inferred from LIGO-Virgo detections so far \citep{abbott2023population}. We use the same resource for the distribution of black hole spin tilts, and the parameters of the beta distribution describing the spin magnitudes, while the spin azimuths are taken to be uniformly distributed. We retain only the aligned components of the spin for further analysis.

We assume that the binaries are isotropically distributed in sky location and orientation, with uniformly distributed coalescence phase and polarization angle. We also assume that the GWs arrive randomly over an 18-month period corresponding to the length of the O4 run.

Our analysis is independent of the normalization of the intrinsic merger rate of BBHs since the pair-wise false alarm probabilities and efficiencies of strong lensing detection are fractional quantities. We \textit{do} need to incorporate the rate of BBH detections when computing the significance of detection in a catalog, and we discuss our choices in \S\ref{sec:catsig}.

\subsection{Lensed BBHs}
Our procedure for simulating strong lensing of GWs is identical to that of \cite{haris2018identifying}, so here we only summarize it and refer the reader to the original paper for details.

We draw lensed BBH sources from the same intrinsic population as described in the previous section, but modify the redshift distribution to also account for the lensing probability $P_\ell(z_s) = 1-e^{-\tau(z_s)}$ determined by the strong lensing optical depth $\tau(z_s)$. We assume these sources are lensed by Singular Isothermal Ellipses (SIE) \citep{kormann1994isothermal, fukugita1991gravitational}, whose velocity dispersion and axis ratio is distributed according to the SDSS catalog of galaxies \citep{collett2015population}, and their redshift is chosen according to the differential optical depth. Finally, the impact parameters are drawn uniformly within the cut/caustic regions of the lens, ensuring that multiple images are obtained. The magnification, time delays, and Morse phases are then calculated by inverting the lens equation numerically.

\subsection{Selection effects}
The end result of the simulations described in the preceding sections is two sets of samples of BBH mergers: an unlensed set, and a lensed set. It is not necessary that each of these simulated BBHs would be detectable by LIGO-Virgo detectors since they are preferentially sensitive to louder signals. To simulate these selection effects, we apply a threshold on the optical signal-to-noise ratio (S/N) computed assuming a noise spectral density representative of the O4 run \citep{H1L1V1-psd-O3O4O5}. An unlensed BBH merger is deemed detectable if it leads to an optimal S/N $\geq 4$ in at least two detectors out of LIGO Livingston, LIGO Hanford, and Virgo, while simultaneously having an optimal network S/N $\geq 8$ in the 3 detector network. For the lensed BBH mergers, we require at least two images to satisfy these criteria.

We find that $\sim 0.78\%$ of the injected population is detectable. Out of the detected events, $\sim 0.15\%$ would be lensed. While intrinsically about 86\% of all lensed images have a Morse phase difference of $\pi/2$, it decreases to about 79\% after applying selection cuts. This means, with reference to eq.~\ref{eq:Morse_phase_factorization}
\begin{eqnarray}
\label{eq:dphi_prior}
P(\dphi=0 \mid \HL=H_L) &=& 0.14 \nonumber \\
P(\dphi=\dfrac{\pi}{2} \mid \HL=H_L) &=& 0.86 \nonumber \\
P(\dphi=0 \mid \HL=H_L^{D}) &=& 0.21 \nonumber  \\
P(\dphi=\dfrac{\pi}{2} \mid \HL=H_L^{D}) &=& 0.79
\end{eqnarray}
We ignore images with Morse phase difference of $\pi$ (i.e. type III images) because they are extremely rare.

\subsection{Parameter estimation}
We draw 1000 samples from the simulated population of detectable unlensed mergers as our background. This number is chosen as a tradeoff between our ability to estimate the search efficiency at low false alarm probabilities, and the computational cost of running such a background analysis. For the foreground, we draw 1000 \textit{pairs} of samples from our simulated population of strongly lensed and detectable image pairs.

We inject each of these 3000 BBH signals into colored Gaussian noise with a power spectral density representative of the O4 run \citep{H1L1V1-psd-O3O4O5}, and perform PE using \texttt{cogwheel} \citep{roulet2022removing, islam2022factorized}. Though the strong lensing detection statistic we discuss in this paper is designed to incorporate all the available information about the BBH signals, we restrict ourselves to quasicircular binaries having spins aligned with the angular momentum and ignore the effects of higher modes of gravitational radiation. This greatly accelerates PE, taking about half an hour for each injection on a single CPU core using the IMRPhenomXAS \citep{pratten2020setting} waveform approximant.

We use uniform priors over detector frame component masses $(m_1^z, m_2^z)$, aligned spin magnitudes $(a_1,a_2)$, polarization angle $\psi$, and geocentric time of arrival $t_c$. We assume isotropic priors on the sky localization (uniform in $\alpha$ and  $\sin{\delta})$, and orientation (uniform in $\cos{\theta_{JN}}$ and $\phi_c$). For the prior on luminosity distance $d_{L}$, we use an empirical fit to the distribution of luminosity distances in our simulated population of detectable unlensed mergers.

\begin{figure}[t]
\centering
\includegraphics[width=\columnwidth]{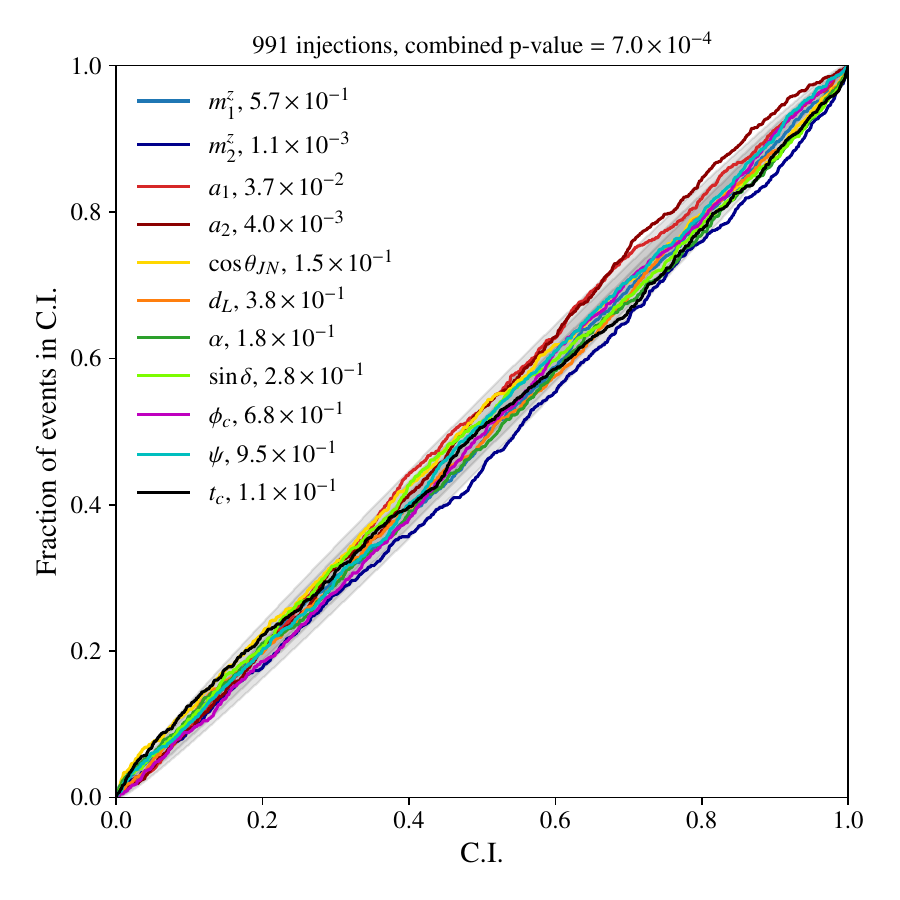}
\caption{This $p-p$ plot shows the fraction $p$\% of the injections whose true value for the parameters lies within the $p$\% credible interval of the estimated posteriors. The plot is computed using 991 PE runs performed using \texttt{cogwheel}. The names of the parameters are shown in the legend, along with their p-value computed using the KS test.}
\label{fig:cogwheel_PP_plots}
\end{figure}

We validate our background PE dataset of 991 successful runs using a $p-p$ plot, which is a frequentist test for Bayesian posterior distributions. If one interprets the Bayesian posterior distributions as the probability distribution of the injected parameter value under different realizations of noise, it is natural to expect that $p$\% of all injections will lie within the $p$\% credible interval, provided that the Bayesian priors were identical to the injected population.

Since our PE priors are different from the population of detectable unlensed mergers, we reweight the posteriors with the correct population prior before computing marginalized credible intervals. In Fig.~\ref{fig:cogwheel_PP_plots}, we show that the fraction of injections recovered within credible intervals is indeed close to the credible interval for all the 11 parameters sampled using \texttt{cogwheel}, as indicated by the high p-values calculated using the KS test. There are slight biases in $m_2$ and spin magnitudes, but we suspect that these are caused by shot noise in prior reweighting, and probably have nothing to do with \texttt{cogwheel}.

\bibliography{po2}

\end{document}